\documentclass[trackchanges,twocolumn]{aastex7}
\usepackage{subcaption}
\begin{document}

\title{Modeling Atmospheric Alteration on Titan: Hydrodynamics and Shock-Induced Chemistry of Meteoroid Entry
}

\author[orcid=0009-0001-9522-8416,sname='Ryushi Miyayama']{Ryushi Miyayama}
\affiliation{Department of Physics, Graduate School of Science, Nagoya University, Furo-cho, Chikusa-ku, Nagoya 464-8602, Japan}
\affiliation{Earth and Planetary Sciences Department, Stanford University, Stanford, CA, USA}
\email[show]{miyayama.ryushi.w1@s.mail.nagoya-u.ac.jp, \\ rmiyayam@stanford.edu}  

\author{Laura Kay Schaefer} 
\affiliation{Earth and Planetary Sciences Department, Stanford University, Stanford, CA, USA}
\email{lkschaef@stanford.edu}

\author{Hiroshi Kobayashi}
\affiliation{Department of Physics, Graduate School of Science, Nagoya University, Furo-cho, Chikusa-ku, Nagoya 464-8602, Japan}
\email{hkobayas@nagoya-u.jp}

\author{Andrea Zorzi}
\affiliation{Earth and Planetary Sciences Department, Stanford University, Stanford, CA, USA}
\email{zorzi@stanford.edu}

\begin{abstract}
Meteoroid entry into planetary atmospheres generates bow shocks, resulting in high-temperature gas conditions that drive chemical reactions. In this paper, we perform three-dimensional hydrodynamic simulations of meteoroid entry using the Athena++ code, {coupled with chemistry calculations via Cantera} to model the non-equilibrium chemistry triggered by atmospheric entry. Our aerodynamical simulations reveal the formation of complex shock structures, including secondary shock waves, which influence the thermodynamic evolution of the gas medium. By tracking thermodynamic parameters along streamlines, we analyze the effects of shock heating and subsequent expansion cooling on chemical reaction pathways. Our results demonstrate that chemical quenching occurs when the cooling timescale surpasses reaction rates, leading to the formation of distinct chemical products that deviate from equilibrium predictions. We show that the efficiency of molecular synthesis depends on the object’s size and velocity, influencing the composition of the post-entry gas mixture. Applying our model to Titan, we demonstrate that organic matter can be synthesized in the present environment of Titan. Also, we find that nitrogen, the dominant atmospheric component, remains stable, while water vapor is efficiently removed, a result inconsistent with equilibrium chemistry assumptions. Moreover, we compare our simulation results with laser experiments and find good agreement in chemical yields. Finally, we also evaluate the impact on Titan's atmosphere as a whole, showing that {meteoroid entry events could have played a significant role in supplying molecules such as HCN during early Titan's history.}
\end{abstract}


\section{Introduction}\label{sec:intro}
Meteorite entry and subsequent impacts play a crucial role in shaping planetary atmospheres and surface conditions \citep[e.g., ][]{abe_matsui1985, abe_matsui1988, chyba_sagan1992}. They contribute to the growth of atmospheres \citep[e.g., ][]{Benlow_Meadows1977,Lange_Ahrens_1982Icar,zahnle+1992}, atmospheric mass loss \citep[e.g., ][]{genda+2005,svetsov2007,shuvalov2009,schlichting+2015,wyatt+2020,kurosaki+2023b}, and chemical processing \citep[e.g., ][]{Lupu+14,zahnle+2020,itcovitz+2022}, which can influence climate evolution and even prebiotic chemistry. Understanding the chemical effects of meteoroid entry and subsequent impact events is therefore essential for reconstructing the atmospheric history of planets and assessing their potential habitability. We have previously studied the chemical composition of impact-generated gases \citep{schaefer+2007,schaefer+2010}, as well as the impact vaporization processes using fluid dynamical simulation \citep{miyayama2024}. Here, we focus on meteoroid atmospheric entry as a precursor to planetary impacts. 

The flow around high-density objects has been extensively studied in various contexts \citep{Landau_fluid,zeldovich+1967}. In astrophysical settings, accretion onto strong gravitational objects, such as black holes and young stars, shares similarities in terms of shock formation and gas dynamics \citep[e.g., ][]{foglizzo+2005, thun+2016, kitajima+2023}. On a planetary scale, meteoroid entry and planetary defense strategies involve analogous shock structures and aerodynamic interactions \citep[e.g., ][]{chyba+1993, shuvalov+2002, Shuvalov+2014, shuvalov+2017}. These studies have shown that such flows generate complex structures. Due to planetary gravitational acceleration, the velocity of impacting objects easily exceeds the sound speed in the surrounding medium, leading to bow shock formation \citep[e.g., ][]{foglizzo+2005, thun+2016, Shuvalov+2014, shuvalov+2017, silber+2017, silber+2018}. This bow shock induces extreme heating of the surrounding gas, which then rapidly cools and decompresses through expansion. Consequently, meteoroid atmospheric entry has the potential to alter atmospheric chemical composition \citep[e.g., ][]{chyba+1990,chyba_sagan1992, ishimaru+2011, silber+2017, flowers+2023}.

Similar to the chemistry induced by atmospheric entry, the chemical composition of impact-generated vapor is governed by shock heating and expansion cooling \citep[e.g., ][]{Mukhin1989, ishimaru+2010, kurosawa+2013}. In shocked vapor, high temperatures accelerate chemical reactions, leading to significant compositional changes. However, as the gas cools, the reaction rate decreases (cf. Arrhenius's law) and eventually becomes slower than the cooling timescale, causing chemical reactions to freeze out \citep[e.g., ][]{fegley+1986,zahnle+1990,gerasimov+1998}. Previous studies have estimated the final chemical products under the assumption of equilibrium \citep[e.g., ][]{hashimoto+2007,schaefer+2007,schaefer+2010, kuwahara+2015}. This assumption is valid if the temperature remains high enough for equilibration and quenching occurs instantaneously at $t = \tau_{\rm cool}$. However, as discussed above, the quenched chemical composition is determined by the interplay of shock heating and subsequent expansion; thus, it is essential to model the time-dependent chemical evolution alongside fluid dynamics.

Studies incorporating non-equilibrium processes via hydrodynamic simulations are limited \citep{ishimaru+2010, ishimaru+2011}. \citet{ishimaru+2010} investigated impact vapor chemistry using a one-dimensional hydrodynamic model under the assumption of spherical symmetric expansion. \citet{ishimaru+2011} simulated shock-driven chemical evolution during atmospheric entry using a one-dimensional steady-state model. However, one-dimensional simulations cannot capture the complex flow structures induced by meteoroid entry, nor can they account for the influence of object size on the flow dynamics. Furthermore, as will be discussed later, the cooling timescale depends on meteorite size, which one-dimensional models fail to incorporate properly, potentially leading to inaccuracies in quench temperature estimation.

In this study, we perform three-dimensional hydrodynamical simulations coupled with chemical kinetic calculations to model the non-equilibrium chemical processes occurring during meteoroid entry into Titan's atmosphere. Titan retains a thick, reducing atmosphere, which is favorable for organic synthesis \citep[e.g., ][]{jones+1987,mckay+1988,scattergood+1989,sekine+2011,ishimaru+2011, flowers+2023, Miller+2025}. To simulate aerodynamic flow, we use the Athena++ code \citep{stone+2020}, which has been widely applied in astrophysical research. Based on the temperature and pressure evolution obtained from these simulations, we conduct chemical reaction modeling using the Cantera chemistry code \citep{goodwin2023}. {In particular, we focus on the production of carbon-bearing species such as HCN, as these molecules are regarded as key precursors to more complex organics, including amino acids.}

This paper is organized as follows. Section~\ref{sec:aerodynamics} presents the aerodynamics component: the setup of the fluid dynamical simulation is described in Sect.~\ref{subsec:athena}, and the results are shown in Sect.~\ref{subsec:dynamics_result}. The chemistry model is addressed in Sect.~\ref{sec:chem}. We describe our simulation methodology in Sect.~\ref{subsec:cantera}, followed by the results of our chemistry simulations in Sect.~\ref{subsec:chem_result}. In Sect.~\ref{sec:discussion}, we discuss atmospheric chemistry induced by meteoroid entry in Titan's atmosphere and a comparison with experimental results and Cassini-measurements (Sect.~\ref{subsec:discuss_titan}), and also discuss the importance of the aerodynamical chemistry compared with the impact-generated chemistry (Sect.~\ref{subsec:vs_impact}). Finally, Sect.~\ref{sec:conclusion} summarizes our findings and conclusions.

\section{Aerodynamics}\label{sec:aerodynamics}
\begin{figure}[htbp]
\centering
   {\includegraphics[width=8cm]{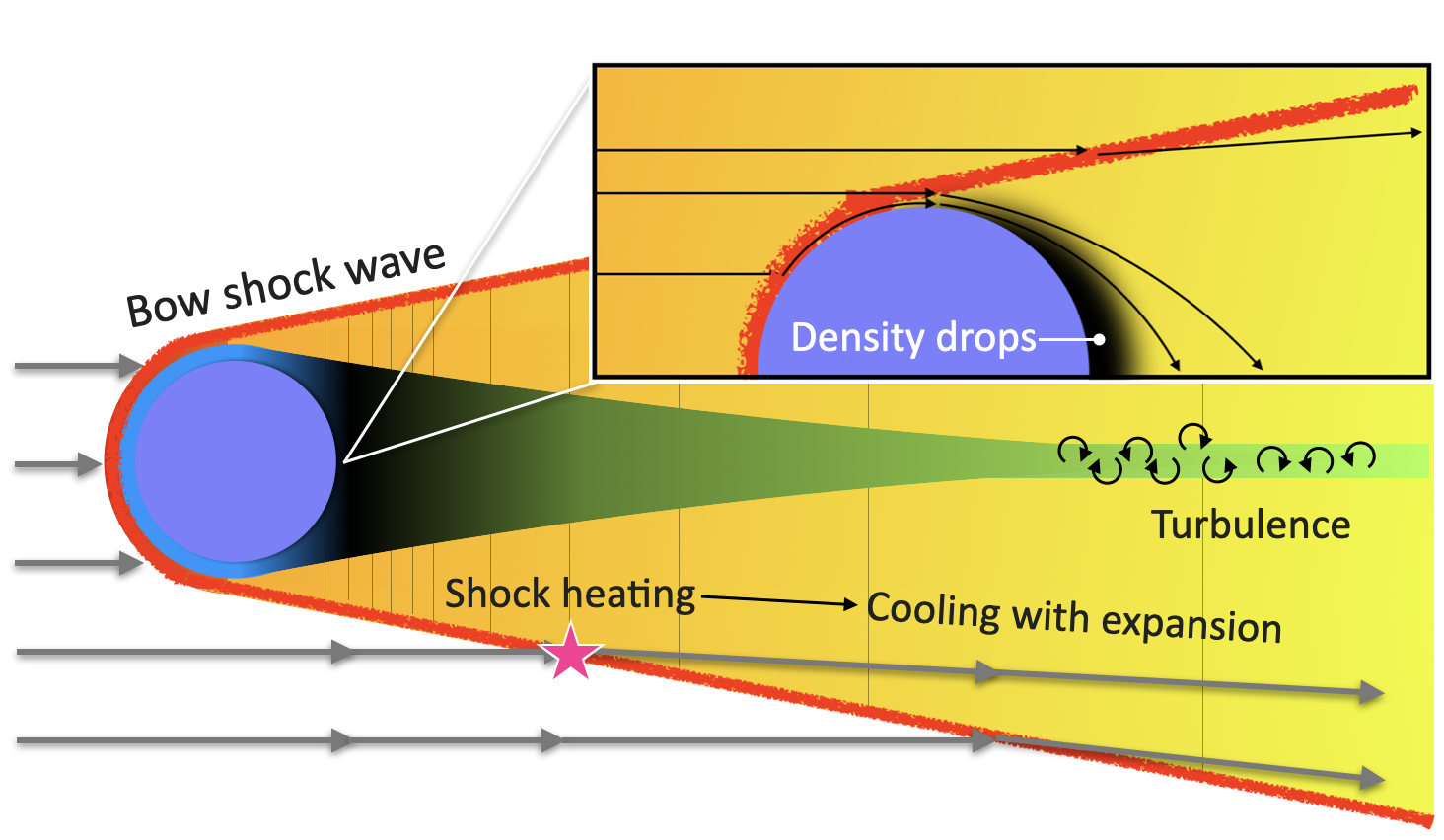}}
\caption{{Schematic depiction of a bow shock flow.} Meteoroid entry results in the bow shock wave (red line). The box illustrates stream lines around the object. Heated medium due to the bow shock immediately cools down with expansion process. Turbulent flow happens at the tail region through a growth of an instability, leading to mixing of material ablated from the projectile with the atmosphere. 
\label{fig:schematic}
}
\end{figure}

Figure~\ref{fig:schematic} shows a schematic depiction of a bow shock flow driven by meteorite entry. A bow shock wave occurs around a meteoroid, leading to atmospheric heating. Chemical reactions due to shock heating processes along a streamline behind the occur bow shock, which is described as gray arrows in Fig.~\ref{fig:schematic}. In this section, we perform hydrodynamical simulations to compute aerodynamical heating through atmospheric entry. Then we show features of the bow shock flow including thermodynamic evolution along a streamline, cooling timescale by expansion, and mixing flow at the wake of the meteorite (Sect.~\ref{subsec:dynamics_result}). 
\subsection{Methods}\label{subsec:athena}
We carry out 3D hydrodynamical simulations using Athena++ code \citep{stone+2020} with the HLLC algorithm as a Riemann solver, which can properly solve contact discontinuity between the projectile and medium. Our simulations are performed in a co-moving frame with the comet and are computed in a cylindrical coordinates $(r,\phi,z)$ based on the symmetry of $\phi$. However, it has been mentioned that the axisymmetric assumption suppresses the growth of  instabilities \citep[e.g., ][]{foglizzo+2005,thun+2016}, which actually develop in the wake of a comet \citep[e.g.,][]{ishimaru+2011} (Fig. \ref{fig:schematic}). In this study, we assume that the cross-sectional areas in the front and rear regions of the comet are small and that the associated chemical reactions can be treated as minor. To simplify the calculations, symmetry in the $\phi$-direction is adopted.

Although our simulation has an azimuthal symmetry $\phi$, Athena++ does not currently support symmetric computation in cylindrical coordinates. Therefore, in all following results, the azimuthal direction is resolved with a coarse resolution of only 2 cells. The computational domain consists of a two-dimensional box with a uniform grid of 600 $\times$ 10000 cells for the $r$ and $z$ directions, respectively. Although impact-induced shock waves are well resolved when using 20 cells per projectile radius (cppr) \citep{miyayama2024}, post-shock expansion is more sensitive to spatial resolution, and 20 cppr results in unphysical errors during the expansion process. Therefore, $50$ cells within the computational box are allocated to simulate the radius of a meteoroid. {In our simulation setup}, high-resolution calculations tend to require a small CFL number. The time step $\Delta t$ is determined by the CFL number $C$, the spatial resolution $\Delta x$, and the characteristic velocity $v$, following the relation $\Delta t = C \Delta x / v$. In our simulations, we use $C = 0.1$, as $C = 0.2$ leads to a computational crash.

{\citet{melosh1981}} showed the critical radius for meteoroid breakup $R_{\rm crit}$ can be described as,
\begin{eqnarray}
R_{\rm crit} = \sqrt{\frac{\rho_{\rm 0}}{\rho_{\rm com}}}H,
\label{eq:r_crit} 
\end{eqnarray}
where $\rho_{\rm com}$ is the meteoroid density, $\rho_0$ is the atmospheric density at sea level and $H$ is the atmospheric scale height. Fragmentation occurs when the radius of an infall body is smaller than $R_{\rm crit}$. We consider the meteoroid entry to Titan's atmosphere for which we assume $\rho_{\rm com}=1.0\times10^3\rm \,kg/m^3$, $\rho_0=1.2\,\rm kg/m^3$ and $H=20\,\rm km$, which yields $R_{\rm crit}=700\,$m. Since we are focusing on atmospheric entry without fragmentation, we assume a spherical object as a comet with the radius of $R_{\rm com}=1\,$km, which is larger than $R_{\rm crit}$. Therefore, the meteoroid mass is $M_{\rm met}=4.0\times10^{12}$\,kg.

\begin{figure*}[htbp]
\centering
   {\includegraphics[width=5.5cm]{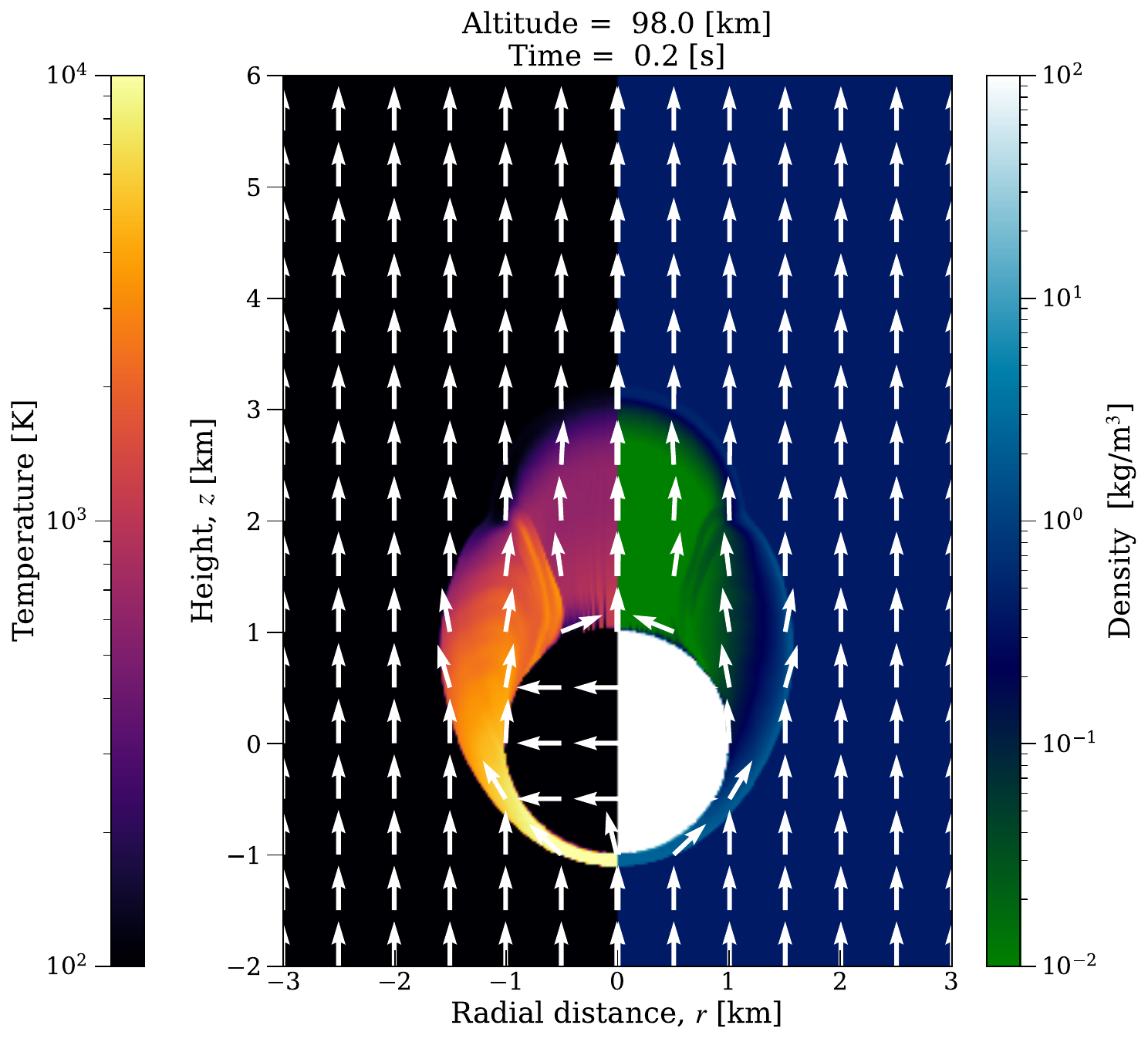}}
   {\includegraphics[width=5.5cm]{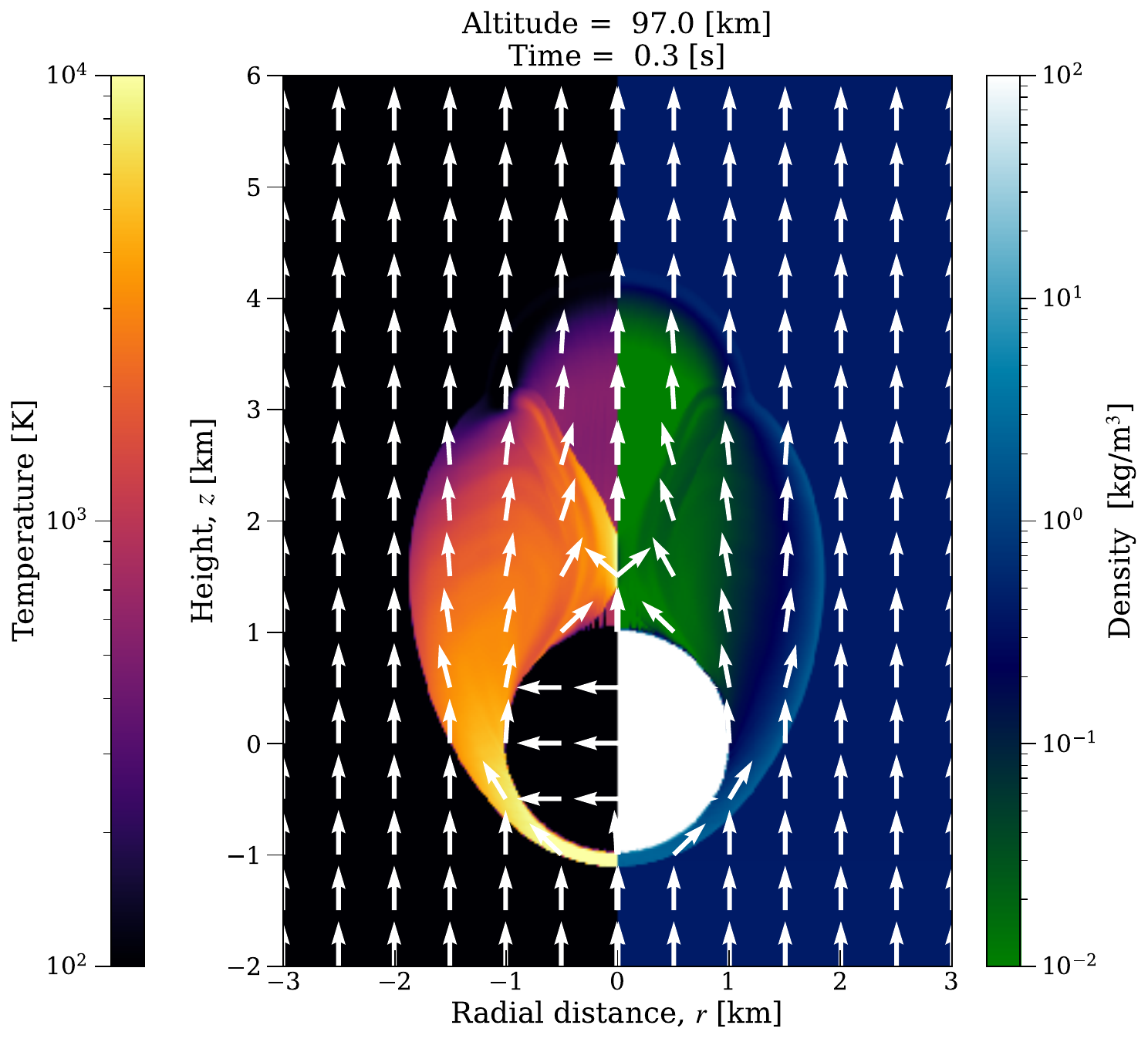}}
   {\includegraphics[width=5.5cm]{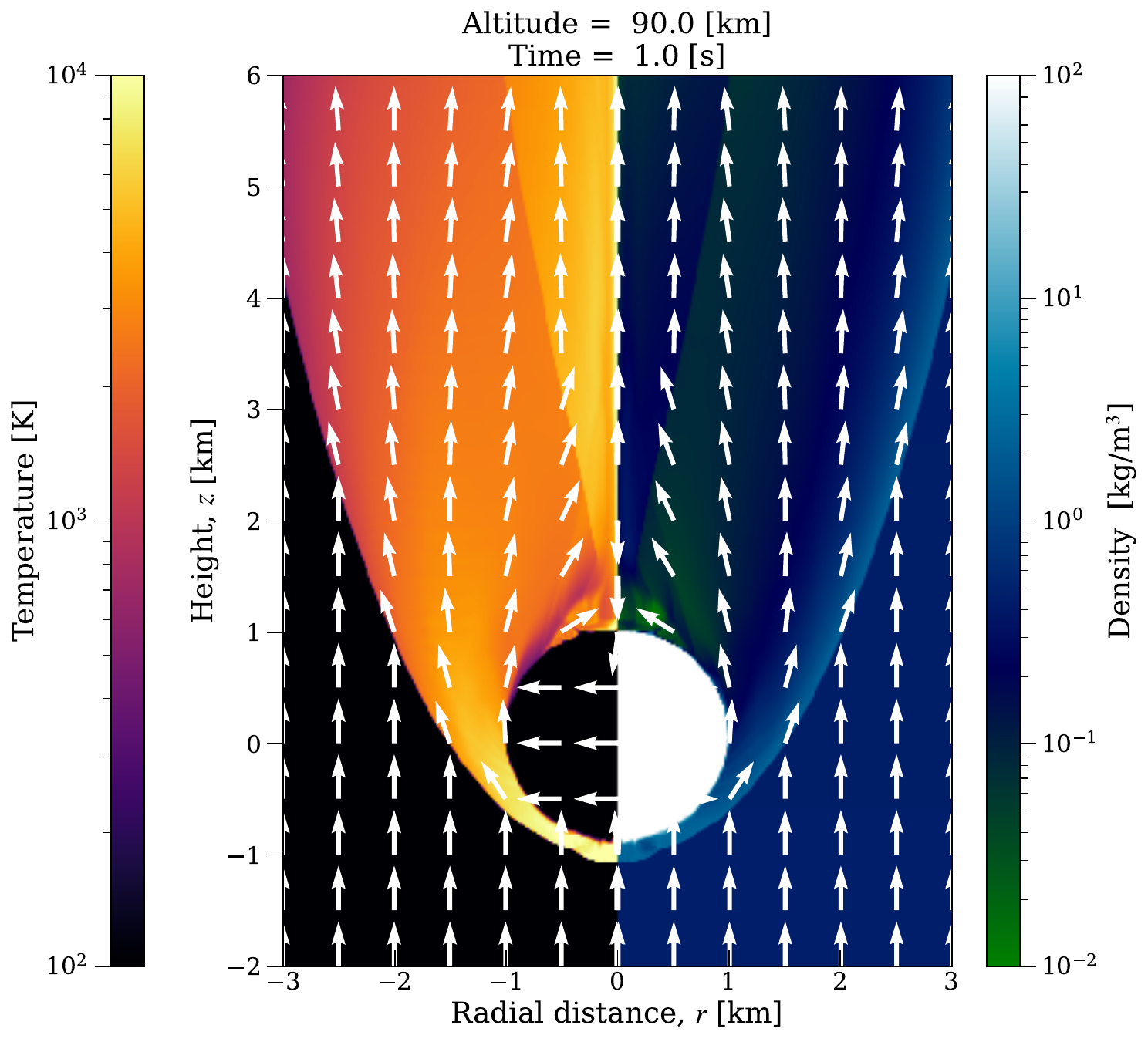}}
\caption{{Overview of the bow-shock formation.} These panels show the zoomed-in views around the cometary body at $t=0.2,\ 0.3$ and $1.0$ s. Colors illustrate the changes in density (right half) and temperature (left half) caused by the bow shock. White arrows indicate the velocity direction, with lengths scaled to 1.
\label{fig:secondshock}
}
\end{figure*}

We assume a meteoroid atmospheric entry velocity $v_{\rm ent}=10$\,km/s. This value is consistent with the theoretical lower limit for $v_{\rm ent}$, which is influenced by the gravitational attraction of both Titan and Saturn. The entry velocity $v_{\rm ent}$ is estimated as, 
\begin{eqnarray}
v_{\rm ent}^2 &\gtrsim& v_{\rm esc, T}^2 +v_{\rm esc, S}^2\\
v_{\rm esc,T} &=& \sqrt{\frac{2GM_{\rm T}}{R_{\rm T}}}\\
v_{\rm esc,S} &=& \sqrt{\frac{2GM_{\rm S}}{a_{\rm T}}}
\label{eq:v_entry} 
\end{eqnarray}
where $v_{\rm esc,T}$ is Titan's escape velocity with $M_{\rm T}=1.3\times10^{23}\,\rm kg$ the mass of Titan, $R_{\rm T}=2.6\times10^3\,\rm km$ the radius of Titan, and $G=6.7\times10^{-11}\,\rm m^3kg^{-1}s^{-2}$ the gravitational constant. $v_{\rm esc,S}$ is the additional term due to gravitational focusing from Saturn's gravity with $M_{\rm S}=5.7\times10^{26}\,\rm kg$ the mass of Saturn and $a_{\rm T}=1.2\times10^6\rm \, km$ the semi-major axis of Titan around Saturn. Therefore, estimated as $v_{\rm ent}\gtrsim8.3\,\rm km/s$, we set the entry velocity $v_{\rm ent}=$ 10\,km/s, {which is consistent with other estimates of the entry velocity on Titan \citep{zahnle03}}. 

We assume that the meteoroid falls into an atmosphere that is well-described by an isothermal hydrostatic-structure described by an ideal-gas equation of state. Thus density and pressure of the medium are given as: 
\begin{eqnarray}
\rho(z)&=&\rho_0 \exp \left(\frac{-mg}{k_{\rm B}T}z \right)
\label{eq:dens_static}\\
P(z)&=&\frac{k_{\rm B}T}{m}\rho(z)=P_0\exp \left(\frac{-mg}{k_{\rm B}T}z \right)
\label{eq:pres_static} 
\end{eqnarray}
where $\rho$, $P$ and $T$ are density, pressure, and temperature, respectively, $m$ is the mean molecular weight, $k_{\rm B}$ is the Boltzmann constant, and $g$ is constant gravitational acceleration. Each quantity with the subscript 0 indicates its value at the surface. We set $\rho_0=1.2\,\rm kg/m^3$, $T=100$\,K and $g=1.4\,\rm m/s^2$ based on the atmospheric structure measurements by the Huygens probe \citep{fulchignoni+2005}, and set the initial altitude of the meteoroid to $100\,\rm km$ $(\simeq5H)$. 

For our simulation, an outflow boundary condition is applied at the outer direction of the $r$ and $z$ axis, while a reflective boundary condition is adopted at the inner side of the $r$-axis, which means the flow going into this direction collides along the $z$-axis. As explained, our frame of computations is fixed with a meteoroid entry, therefore the density and pressure of the incoming flow from the bottom increase as time passes to account for atmospheric hydrostatics. 

When a fluid strikes a finite object, the density and pressure directly behind the object become extremely low \citep{Landau_fluid}. This significant drop of density and pressure in the wake of the object occasionally leads to numerical instabilities. To mitigate this issue, we introduce floor values for density ($10^{-6}\,\rm kg/m^3$) and pressure ($10^{-6}$ Pa), which are over 3 orders of magnitude smaller than the initial density and pressure used in the calculations.

All simulations for atmospheric entry are represented by ideal hydrodynamics without any diffusion processes. Meteoroid fragmentation and ablation are driven by hydrodynamical drag-force, rather than conduction or radiative process. Bow-shock flow results in a pressure difference across a meteoroid body: the front of a meteor faces high pressure, while the rear experiences extremely low pressure. The drag force from this differential pressure causes atmospheric breakup of the impactor \citep[e.g., ][]{passey+1980,melosh1981,chyba+1993}. Also, heat conduction and radiation processes are neglected because their timescales are long compared to {the chemical-quenching timescale}. At the scale of a few meters scale, which is the minimum length resolved in our computations, the timescales of heat conduction and radiation are estimated to be $\tau_{\rm heat}\sim10^6 \rm\,s$ and $\tau_{\rm rad}\sim10^2\rm\,s$, respectively (see Appendix~\ref{appendix2}). Compared with the chemical-quenching timescale ($\sim10\,$s), {as discussed in Sect.~\ref{subsubsec:quench}}, those are sufficiently long to be ignored. 

Finally, we describe the treatment of the meteoroid body in our simulations. The atmospheric flow and the meteoroid entry are primarily governed by the large density contrast (at least $10^3$) between the projectile (1000 kg/m$^3$) and the atmosphere (1.2 kg/m$^3$), rather than the equation of state (EoS) of the projectile. Therefore, since we are interested in the chemistry generated by the shock wave from the meteor's passage, rather than the chemistry of the impactor itself, we adopt the ideal gas EoS to describe the meteoroid thermodynamical response to the flow:  
\begin{eqnarray}
P = (\gamma-1)\rho E
\label{eq:eos} 
\end{eqnarray}
where $E$ and $\gamma$ are the internal energy and the heat capacity ratio, respectively. Assuming the atmosphere is made largely of diatomic molecules, we set to $\gamma=1.4$ for both the medium and meteor. We discuss more detailed validation of this EoS assumption in Appendix~\ref{appendix1}. 

\subsection{Results}\label{subsec:dynamics_result}
\subsubsection{Thermodynamics in bow-shock flow}\label{subsubsec:track}
\begin{figure*}[htbp]
\centering
   {\includegraphics[width=16cm]{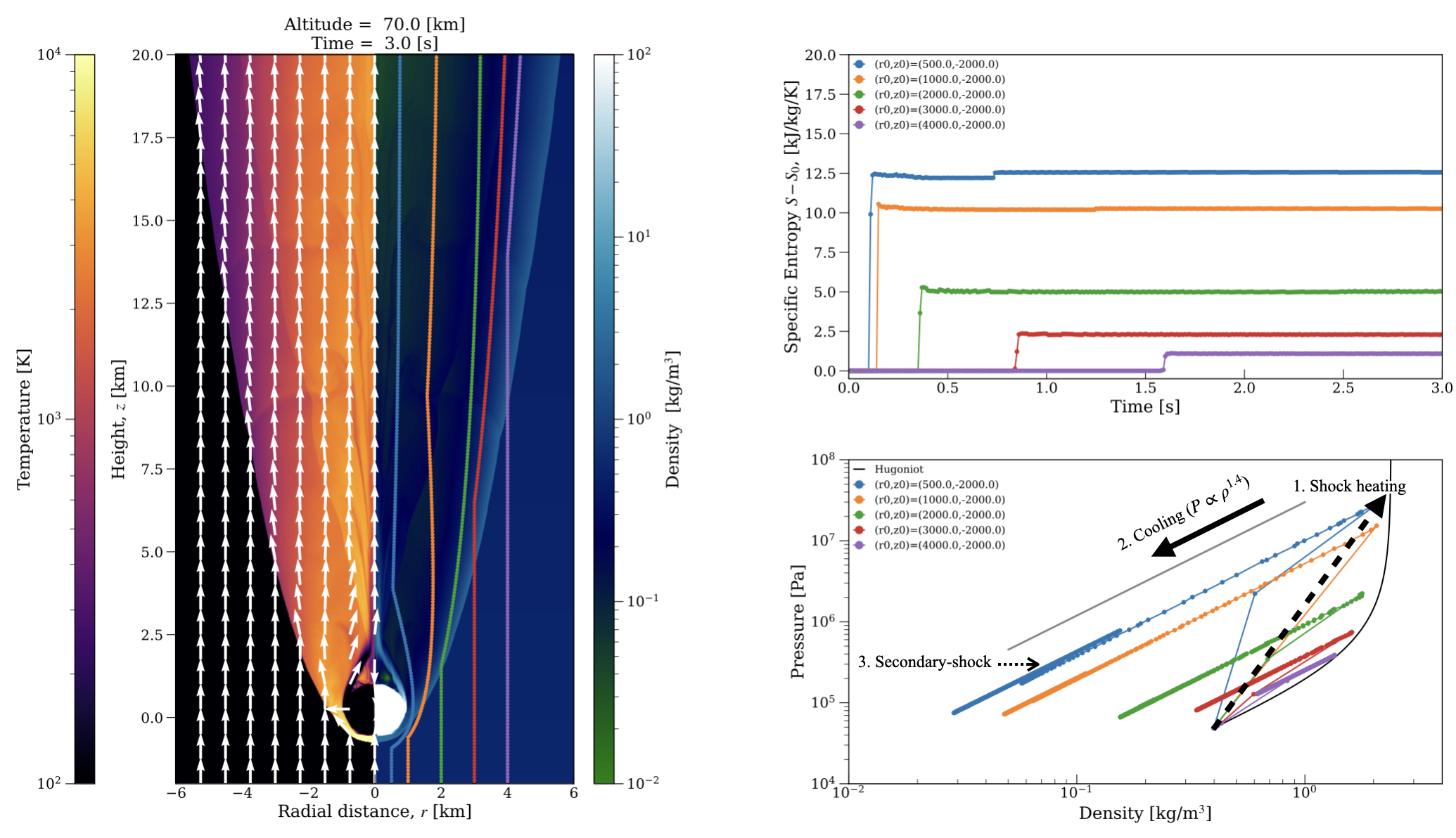}}
\caption{{Tracking particles in bow-shock flow.} In the left panel, the bow shock is illustrated in temperature and density with trajectories of each test particle included, each of which has an individual initial position. In the upper right panel, entropy increases for each particle are described as a function of time. In the bottom panel, the evolution of density and pressure are shown. The Gray curve and line are Hugoniot curves, which represent all possible states after shock heating, and are a function of $P\propto\rho^{1.4}$, respectively. Each particle has a different initial position, and those colors are common in these three panels. 
\label{fig:track}
}
\end{figure*}

Figure \ref{fig:secondshock} displays the zoomed-in views around the comet at $t=0.2,\ 0.3,$ and $1.0\,\rm s$, and shows the bow-shock generation. Supersonic atmospheric-flow collides with the comet body. Then, the bow shock wave forms at $t=0.2\,\rm s$ and heats and compresses the atmosphere in front of the comet, while density becomes extremely small at the rear (the left panel). The atmospheric flow in front of the comet wraps around to the rear, where it converges (the center panel). This converging flow generates a secondary shock wave inside the bow shock behind the comet (the right panel).

Next, we discuss the bow-shock flow in greater detail from a thermodynamic perspective, using the test particle analysis. In Fig.~\ref{fig:track}, three panels are shown--the bow-shock flow with color-maps of density and temperature, entropy of each particle as a function of time, and density-pressure evolution. The left panel illustrates density and temperature with color maps, and also shows trajectories of test particles and velocity vectors (white arrows). As explained in Sect.~\ref{subsec:athena}, our simulations are performed based on a co-moving frame with the meteoroid body, so atmospheric flow is moving from the bottom towards the top of the panel. As also shown in Fig.~\ref{fig:schematic}, particle trajectories are perturbed by the passage of the shock wave. 

The upper right panel of Fig.~\ref{fig:track} shows entropy changes from the initial entropy as a function of time. Entropy jumps as the particle is shocked. For example, the red particle that is initially positioned at $(r_0,z_0)=(3000\,\rm m,-2000\,m)$ experiences an entropy jump at $t=0.8\,\rm s$. Then, the entropy does not significantly change after the entropy jump due to the shock heating. This shows that the expansion process after shock heating is adiabatic. Particles with small $R_0$, i.e., initially closer to the meteoroid symmetry axis, experience strong shocks. The maximum entropy, temperature and pressure decrease with increasing $R_0$. The entropy of the blue line increases again during the expansion phase after the passing bow shock wave. This indicates, as also shown in Fig.~\ref{fig:secondshock}, the secondary shock which is generated behind the moving object. A similar shock structure was found in a simulation of a gravitational accretion process, in which it was found that the object size can influence the flow behind the bow shock \citep{prust+2024}. Impact of this secondary heating on chemistry is discussed in Sect.~\ref{subsubsec:quench}. 

The bottom panel of Fig.~\ref{fig:track} illustrates thermodynamic changes in density and pressure. The gray curve corresponds to Hugoniot curve while the straight gray line is Poisson's law $P\propto\rho^{1.4}$. Hugoniot curve is derived from three conservation laws across the wave (mass, momentum and energy), and {shocked states should discontinuously jump to a point on the curve described as the black-dashed arrow in Fig.~\ref{fig:track}. In a post-shocked state, pressure and density continuously decrease through adiabatic expansion, which is illustrated as the black-solid arrow following the adiabatic expansion $P\propto\rho^{1.4}$. Pressure of the blue particle, initially located at $(r_0,z_0)=(500\,\rm m,-2000\,m)$, for example, is discontinuously compressed to around $10^7$\,Pa due to the passage of the shock wave.} Our simulation well represents post-shock states as expected by Hugoniot curve and Poisson's law. The temperature and pressure evolution found in the hydrodynamical simulation are used as input for the chemical part.

\subsubsection{Cooling time scale}\label{subsubsec:timescale}
As discussed in Sect.~\ref{subsubsec:track}, the gas cools through adiabatic expansion after being heated by the bow shock. As described in Sect.~\ref{sec:intro}, chemical reaction rates decrease as the gas cools, and eventually, the reactions quench. Therefore, the cooling timescale plays a crucial role in determining the final chemical composition. In this section, we discuss the cooling timescale due to adiabatic expansion.

\begin{figure}[htbp]
\centering
   {\includegraphics[width=8cm]{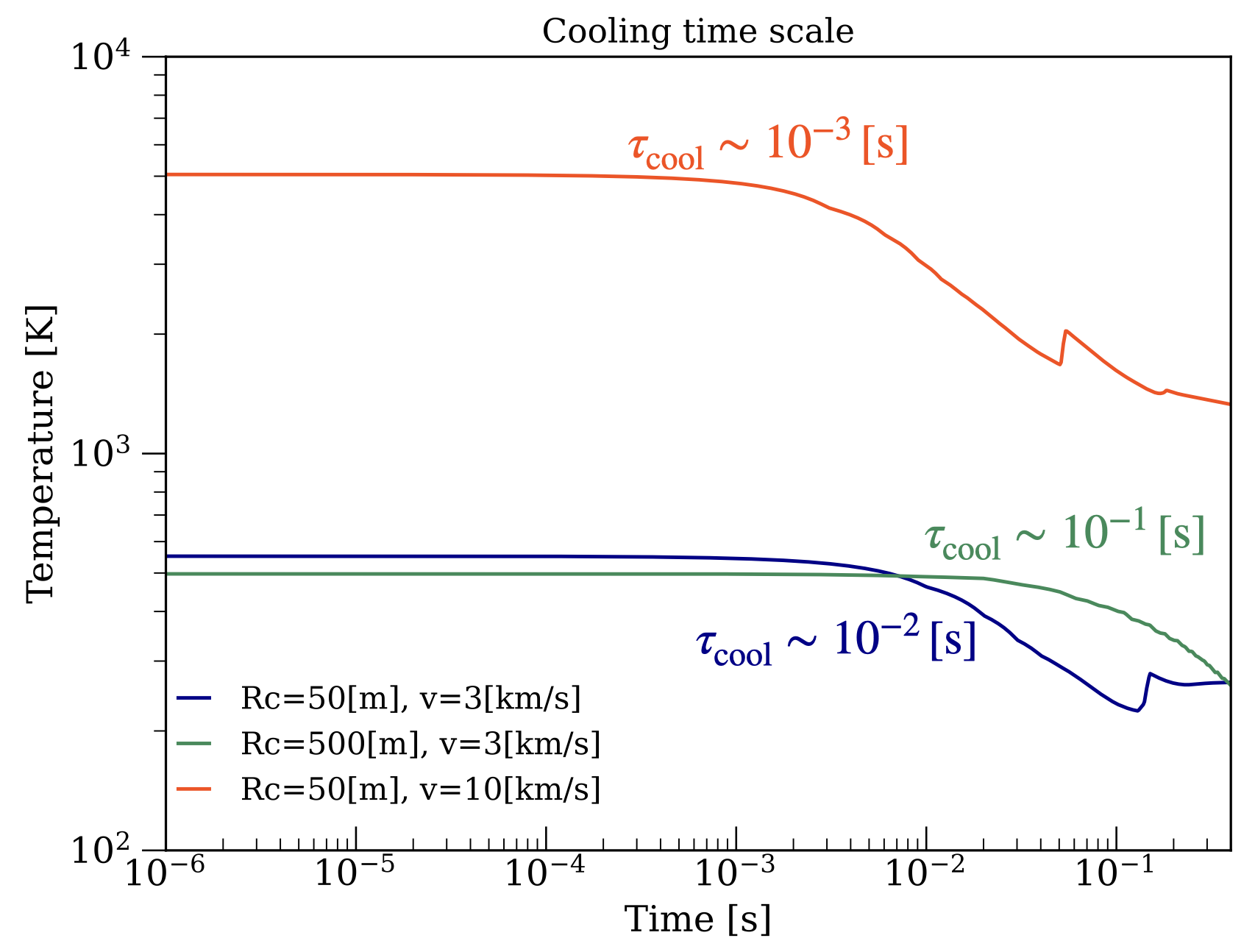}}
\caption{{Cooling time scales via adiabatic expansion.} A line illustrates the temperature of a test particle with an initial position $R_0$ at each time. These colors shows three different simulations. As for the labels in this picture, $R_{\rm c}$ and $v$ is the comet radius and velocity. In these lines, the initial distances from the comet $R_0$ are set to $1.1R_{\rm c}$, e.g., $R_0=55\rm \,m$ in the red-line case. 
\label{fig:cooling}
}
\end{figure}

In our simulations, the characteristic length and velocity of the flow are determined solely by the meteoroid size and entry velocity. Therefore, the cooling timescale, which is the time required for the temperature to decline below an initial nearly isothermal value, can be expressed as follows: 
\begin{eqnarray}
\tau_{\rm cool}\sim R_{\rm com}/v_{\rm ent}.
\label{eq:cooling} 
\end{eqnarray}
where $R_{\rm com}$ is the radius of the comet. To empirically validate this estimation, we compute three different simulations with parameters of entry velocities and sizes. We note that, in only this section, the meteoroid size and velocity are different from Sect.~\ref{subsec:athena}. Figure~\ref{fig:cooling} shows temperature changes as a function of time, which corresponds to a test particle initially located at $1.1R_0$ where $R_0$ is the initial distance from the meteoroid. According to Fig.~\ref{fig:cooling}, Eq.~\ref{eq:cooling} reproduces the cooling time scale well. A minor but important detail in the calculations is that the data output interval for thermodynamic evolution (e.g., Fig.~\ref{fig:track}) must be set finely enough to resolve this timescale. Influence of the cooling timescale on chemistry is discussed in Sect.~\ref{subsubsec:noneq}

\subsubsection{Mixing flow in the tail region}\label{subsubsec:mixing}
The Athena++ code currently supports passive scalar transport analysis. A passive scalar follows the fluid but does not influence the fluid behavior (e.g., ink motion). This analysis enables us to simulate the transport of cometary components by solving a diffusion equation: 
\begin{eqnarray}
    \frac{\partial \rho C}{\partial t}+\vec{\nabla}\cdot (\rho C\vec{v})=0
    \label{eq:passive}
\end{eqnarray}
where $\rho$, $\vec{v}$ are the fluid density and velocity, and $C$ is a passive scalar, which corresponds to the mixing fraction of the atmosphere with the cometary material in our simulation. Although the diffusion term originally appears on the right-hand side of Eq.~\ref{eq:passive}, we set the diffusion coefficient to zero. This is because the diffusion time scale of vaporized gases from the comet into the atmosphere is extremely long, making its effect negligible. While the exact timescale depends on the medium, it generally exceeds $10^4$ seconds for meter-sized objects, whereas our simulation focuses on dynamics over approximately 10 seconds. Therefore, diffusion can be safely ignored.

\begin{figure}[htbp]
\centering
   {\includegraphics[width=8cm]{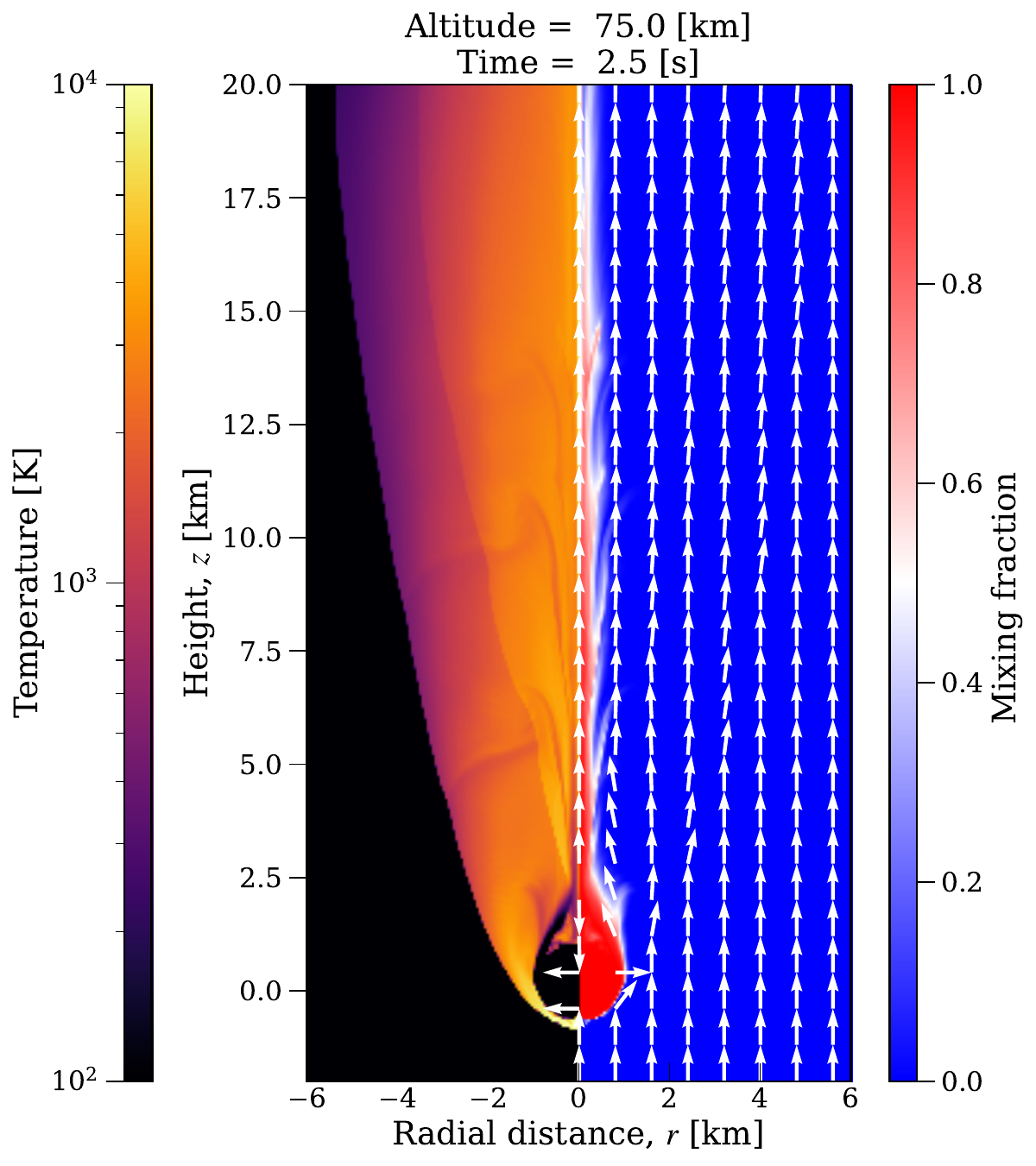}}
\caption{{Temperature and mixing ratio of cometary and atmospheric components.} In the right side ($r>0$), red represents comet-derived components, while blue indicates atmospheric components. 
\label{fig:passive}
}
\end{figure}

Figure~\ref{fig:passive} illustrates the mixing fraction of cometary and atmospheric material. As shown in Fig.~\ref{fig:schematic}, the cometary material is carried by the atmospheric flow toward the rear of the comet rather than in the direction of the shock wave (Fig.~\ref{fig:passive}). In following chemistry simulations, we assume that the cross-sectional area of this region is small and thus neglect the contribution of mixing. However, as discussed in Sect.~\ref{subsec:athena}, non-axisymmetric simulations are expected to exhibit instability growth in the mixing region behind the comet \citep[e.g., ][]{foglizzo+2005,thun+2016}, further enhancing the mixing process. In addition, the temperature in the wake becomes sufficiently high to trigger chemical reactions (Fig.~\ref{fig:passive}). As a result, this mixing region may contribute to the formation of unique chemical species, warranting further consideration in future studies.

\section{Aerodynamical chemistry}\label{sec:chem}
\subsection{Methods}
\label{subsec:cantera}
To simulate chemical reactions induced by aerodynamical heating, {we use the chemical software Cantera, an open-source software package for chemical kinetics, thermodynamics, and transport modeling \citep{goodwin2023}}. Cantera is a powerful tool for modeling non-equilibrium chemical kinetics, thermodynamic properties, and combustion processes. It allows for the integration of time-dependent chemical evolution, making it well-suited for studying the transient conditions induced by meteoroid entry. Compared to equilibrium-based approaches, Cantera enables us to track reaction pathways under dynamically evolving temperature and pressure conditions. We consider gas-phase reactions in the $\rm C/H/N/O$ system using the chemical network, the GRI Mech 3.0\footnote{\url{http://combustion.berkeley.edu/gri-mech/releases.html}}, which is provided in Cantera.  

The temperature and pressure inputs for these chemical simulations are derived from hydrodynamical simulations (Sect.~\ref{subsec:athena}) via the test particle analysis (see Fig.~\ref{fig:track}). This approach tracks the thermodynamic history of a selected parcel of gas, allowing us to capture the thermal and chemical evolution over time. The assumed chemical network does not include ionization processes and provides a parameterization of thermodynamical properties for species up to 6000 K: at higher temperatures, the network extrapolates properties that are unreliable. Such high temperatures are experienced at the head of the meteoroid, where gas can cool radiatively. However, the radiative cooling timescale is two orders of magnitude longer than adiabatic cooling (Appendix ~\ref{appendix2}). Moreover, at such high temperatures, reaction rates become extremely fast, rapidly driving the gas mixture toward chemical equilibrium. Therefore, to model chemistry in these high-temperature regions, we assume adiabatic expansion while maintaining equilibrium until the temperature drops below 6000\,K. Once below this threshold, we switch to a non-equilibrium chemistry simulation.

Titan has a thick, reducing atmosphere and has been studied as a fertile environment for the formation of organic compounds and prebiotic molecules \citep[e.g., ][]{sekine+2011,ishimaru+2011,flowers+2023,Miller+2025}. At present, Titan possesses a massive $\rm N_2$ atmosphere with $\sim 6$ \% volume mixing ratio of $\rm CH_4$ \citep{strobel2010,horst2017}. Cassini-Huygens observations suggested that $\rm N_2$ on Titan is a non-primordial component \citep[e.g., ][]{niemann+2005}. It has been confirmed that $\rm N_2$ on Titan can be produced through meteoroid impact process, both experimentally \citep{sekine+2011} and numerically \citep{ishimaru+2011}. Therefore, we assume a purely gaseous atmosphere dominated by primarily $\rm N_2$ (90\%) with 10\% $\rm CH_4$. Although some solid components, such as dusts and hazes, have been expected to be present \citep{danielson+1973}, {we do not take them into consideration because we do not have reaction rate coefficients for organic haze chemistry.}

The simulation setup is the same as described in Sect.~\ref{subsec:athena}, considering a meteoroid falling into Titan's atmosphere with the density $\rho_0=1.2$\,kg/m$^3$ from an altitude of 100 km with a velocity of 10 km/s. However, unlike the aerodynamical simulations, the initial time ($t=0$) is set to the time when the particles is shocked, which means $t$ is defined as the time after shocked. 
We assume that $\rm N_2$ is already present in the atmosphere before atmospheric entry of the meteorite. Therefore, the atmosphere is a binary mixture of $\rm N_2$ and $\rm CH_4$ with their compositional ratios treated as parameters. We note that the results presented in Sect.~\ref{subsec:chem_result} only include species containing C, H and N. O-bearing species are not considered in this section. CHNO chemistry is discussed in the following section (Sect.~\ref{subsec:discuss_titan}). 

\subsection{Results}\label{subsec:chem_result}
In this section, we show results of chemistry simulations. As explained in Sect.~\ref{subsec:athena}, we assume an atmosphere described by an isothermal hydrostatic-structure. Therefore, while density and pressure vary with altitude, the impact of these variations on chemical evolutions is minor (see Appendix~\ref{appendix3}). Hence, all the results presented below are derived from the test-particle analysis which is located at an altitude of 100 km.

\subsubsection{Quench temperature and secondary shock}\label{subsubsec:quench}
\begin{figure*}[htbp]
\centering
    \begin{subfigure}[t]{0.45\textwidth}
      \caption{$r_0 = 1300$\,m} 
    \includegraphics[width=\textwidth]{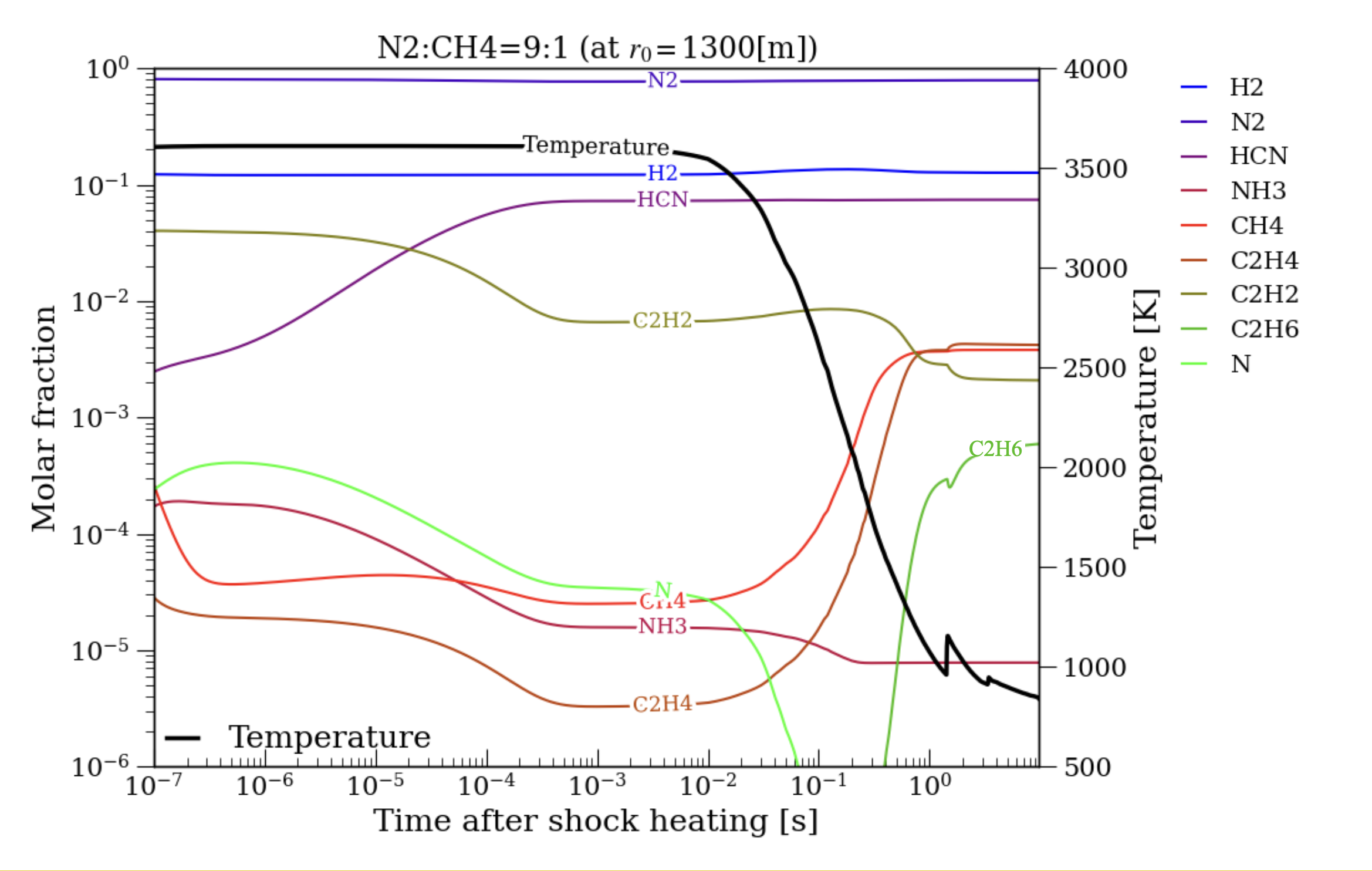}\label{fig:1300}
    \end{subfigure}
    \hfill
    \begin{subfigure}[t]{0.45\textwidth}
      \caption{$r_0 = 1700$\,m}
      \includegraphics[width=\textwidth]{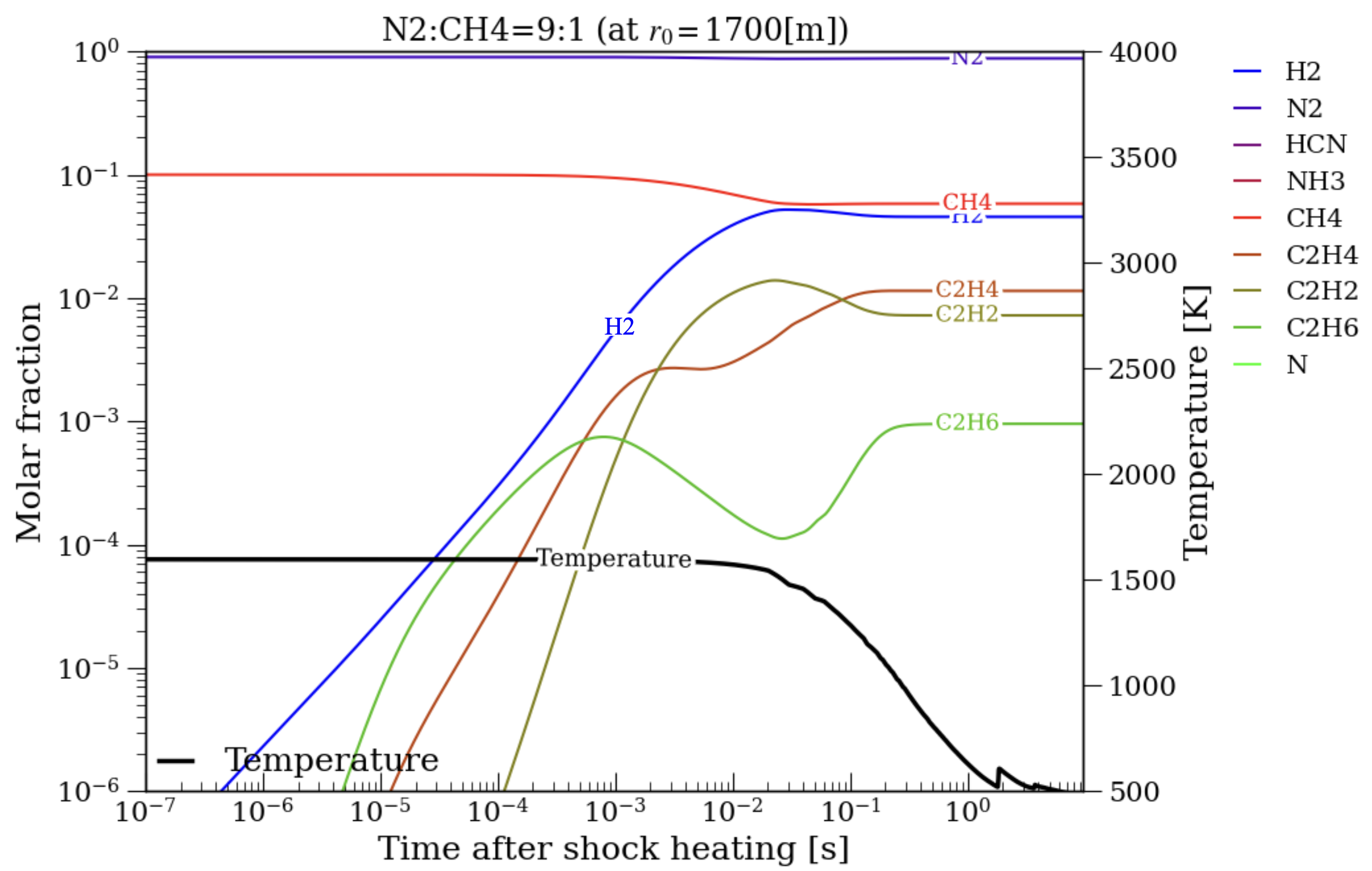}\label{fig:1700}
    \end{subfigure}
   \caption{{Time evolution of chemical species after shock heating.} In this figure, the density of atmosphere, the entry velocity, and the altitude are set to $\rm 1.2\,kg/m^3,\ 10\,km/s$ and $\rm 100\,km$, respectively. Each color shows different chemical components, while the black curve displays the temperature change. The difference between panels is the initial position of the test particle: (a) $r_0=1300$\,m and (b) $r_0=1700\,$m. The temperature increase at $\sim 1\,$s is caused by the secondary shock wave (see Sect.~\ref{subsubsec:track}).
\label{fig:distance}
}
\end{figure*}
First, we present basic results of the chemistry evolution. As shown in Fig.~\ref{fig:track}, the maximum temperature a test particle experiences depends on the radial distance from the comet: since the shock wave becomes stronger when closer to the comet, particles trajectories closer to the comet reach higher maximum temperatures. Therefore, the positional dependence influences chemistry.

Figure~\ref{fig:distance} shows the results of two chemistry simulations with different initial distances of $r_0=1300\rm \, m(a),\ and\ 1700\,m$(b), where $r_0$ is an initial radial position. The plots shows the molar fractions of each species as a function of time after shock heating. The pre-shock atmospheric composition is assumed to be $\rm N_2:CH_4=9:1$. In Fig.~\ref{fig:1300}, non-initial species, such as $\rm H_2$ and $\rm C_2H_2$, appear at the beginning of the figure at $10^{-7}$\,s. This is because the shock temperature is high enough to rapidly produce these species from $\rm N_2$ and $\rm CH_4$ within this short timescale. 

As mentioned above, the temperature (right $y$-axis) in the left panel (Fig.~\ref{fig:1300}) is higher than that in the right panel (Fig.~\ref{fig:1700}). Comparing the two cases, polyatomic species are favored at low temperature, which is consistent with \citet{ishimaru+2010}. In Figure~\ref{fig:1700}, N-bearing molecules are not produced. This is because the temperature at $r_0=1700$\,m is not elevated enough to destroy the N$_2$ bond.

Reaction rates become slow as temperature drops, and then chemical quenching occurs when the chemistry time scale starts to balance with the cooling time scale \citep[e.g., see ][]{fegley+1986}. In impact-induced vapors, gas composition has been found to quench at 1000--3000 K \citep[e.g., ][]{Mukhin1989,gerasimov+1998,ishimaru+2010}. Seeing Fig.~\ref{fig:distance}, chemical quenching happens at $\sim$1\,s in an aerodynamical-heated atmosphere, which corresponds to 500--1000\,K. 

As we discussed in Sect.~\ref{subsubsec:track}, the secondary shock wave that forms behind the bow shock (Fig.~\ref{fig:secondshock}) was not predicted in previous studies \citep{ishimaru+2011}. We discuss the influence of the secondary shock wave on the chemical evolution of the atmosphere. As shown in Fig.~\ref{fig:1300}, the test particle with the initial distance of $r_0=1300$\,m experiences secondary shock heating at 1\,s. 

The impact of the secondary shock on molecular composition varies depending on the species. In particular, C$_2$H$_6$ (ethane) and C$_2$H$_4$ (ethylene) exhibit changes due to the secondary shock heating, whereas CH$_4$ (methane) does not show a variation. 
Furthermore, species that have already undergone chemical quenching, such as HCN, remain unaffected by the secondary shock wave. Since these molecules have reached a kinetically stable state before experiencing the second shock, their abundances do not change significantly, even under the transient temperature and pressure fluctuations induced by the shock passage. Although secondary shock heating also happens at 1\,s in Fig.~\ref{fig:1700}, the fraction changes have already quenched at $10^{-1}$\,s resulting in no-effect from the secondary shock. At larger distances from the comet $r_0$, the shock temperature becomes lower, leading to quickly falling below the quenching temperature ($\sim$ 1000\,K). As a result, the influence of the secondary shock becomes less significant at greater distances from the comet. 

\subsubsection{Equilibrium vs non-equilibrium}\label{subsubsec:noneq}
Atmospheric chemistry has been investigated mainly in the context of planetary impacts, based on the chemical equilibrium assumption \citep[e.g., ][]{hashimoto+2007,schaefer+2007,schaefer+2010,kuwahara+2015}. This assumption would be valid if the shock temperature is sufficiently high to equilibrate the heated atmosphere. {However, as discussed in Sect.~\ref{subsec:athena}, the rapid temperature drop due to adiabatic expansion after shock heating likely invalidates the chemical equilibrium assumption.} \citet{ishimaru+2010} reported that chemical equilibrium calculations underestimate $\rm HCN$ production by an order of magnitude, based on chemistry simulations coupled with 1D hydrodynamical simulations. We discuss how this difference influences chemical composition. 

\begin{figure}[htbp]
\centering
   {\includegraphics[width=8.5cm]{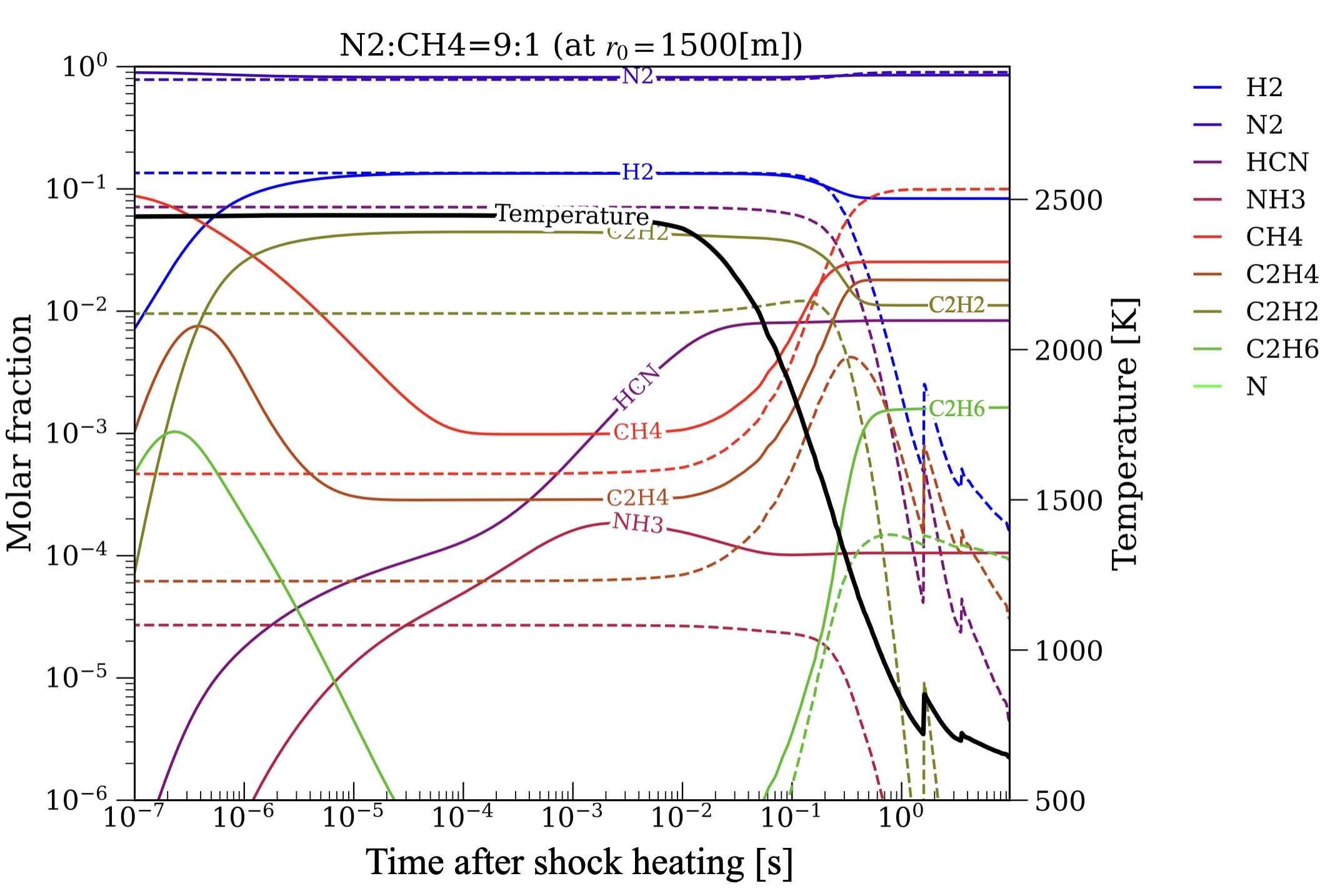}}
\caption{{Comparison of chemical equilibrium and non-equilibrium calculations.} The solid line corresponds to the non-equilibrium calculation, while the dashed line represents the equilibrium calculation. The black line indicates temperature, corresponding to the right axis. This figure is for a simulation with the same impact parameters (density, velocity, and altitude) as in Fig.~\ref{fig:distance}, but with a different initial location ($r_0$=1500\,m).
\label{fig:eq_non1}
}
\end{figure}

Figure~\ref{fig:eq_non1} illustrates two results of chemical evolution from different simulations: equilibrium and non-equilibrium. Solid lines correspond to non-equilibrium case, and equilibrium case is represented as dashed lines. In the equilibrium calculations, chemical abundances are assumed to reach equilibrium instantaneously at every time step, representing the composition that would be achieved if all reactions proceeded to completion immediately. 

As shown in Fig.~\ref{fig:eq_non1}, before the cooling begins ($t\lesssim 10^{-2}\,\rm s$), the fraction of $\rm H_2$ (blue solid-line) reaches its equilibrium abundance at $~10^{-5}$\,s. However, as the temperature decreases due to expansion, the fraction difference of $\rm H_2$ between equilibrium and non-equilibrium simulations becomes apparent. Eventually, in non-equilibrium simulation, chemical reactions quench, resulting in a chemical composition that deviates from equilibrium. 

In contrast, other species, such as $\rm NH_3$ and $\rm CH_4$, do not reach equilibrium before expansion cooling starts. Consequently, their final abundances are determined without ever experiencing an equilibrium state. This occurs because chemical reactions do not progress significantly within the timescale at which cooling becomes effective, as shown in Fig.~\ref{fig:cooling}. At positions farther from the comet, where the shock wave temperature does not become high enough, cooling begins before equilibrium is reached, leading to a composition that can significantly deviate from equilibrium. {As shown in Fig.~\ref{fig:eq_non1}, the predicted HCN abundance differs by approximately three orders of magnitude between equilibrium and non-equilibrium calculations, with molar fractions of about $4\times10^{-6}$ and $1\times10^{-2}$ at $t\simeq10\,$ s, respectively. This result highlights the critical role of non-equilibrium chemistry in enhancing HCN production under realistic post-shock conditions. We will discuss the total contribution to HCN production by a meteoroid entry event in Sect.~\ref{subsubsec:hcn}.}

\section{Discussion}\label{sec:discussion}
\subsection{Formation of Titan's atmosphere}\label{subsec:discuss_titan}
\subsubsection{Overview}\label{subsubsuc:overview}
\begin{figure}[htbp]
\centering
   {\includegraphics[width=8.5cm]{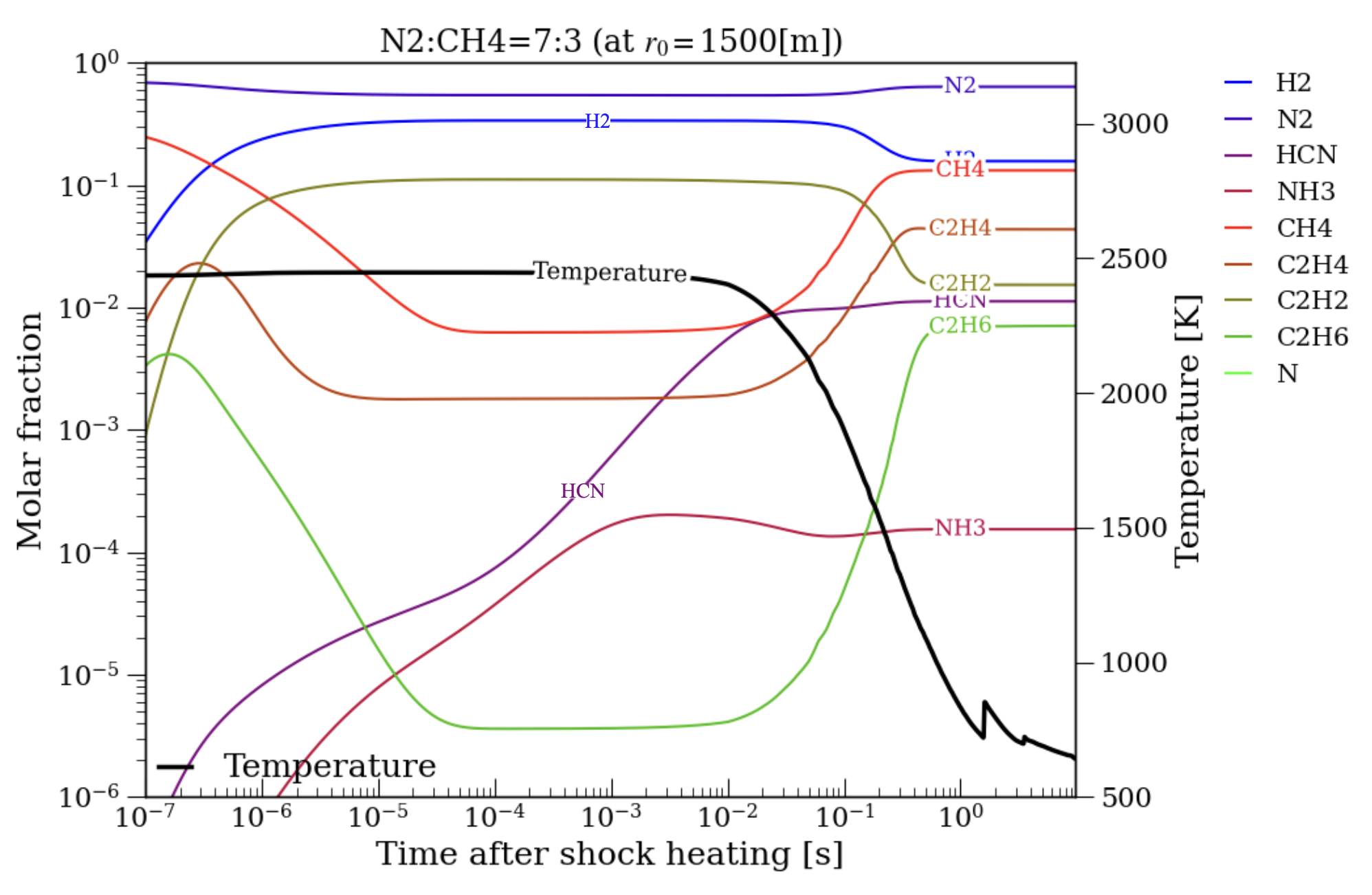}}
\caption{{Dependence of initial fraction of chemical species.} This figure is for a simulation with the same impact parameters (density, velocity, and altitude) as in Fig.~\ref{fig:distance}, but with a different initial nitrogen-to-methane ratio. The initial position of the test particle is $r_0=1500$\,m.  
\label{fig:nitrogen}
}
\end{figure}

{Figure~\ref{fig:nitrogen} presents the results of simulations assuming an atmosphere with a different initial ratio of $\rm N_2:\rm CH_4=7:3$.} Comparing Fig.~\ref{fig:nitrogen} with Fig~\ref{fig:eq_non1}, we find that carbon-bearing species, such as $\rm CH_4,\,C_2H_6$, are strongly influenced by the initial compositional ratio. On the other hand, nitrogen-bearing species, specifically $\rm HCN$, do not change with different initial atmospheric composition. This indicates that the nitrogen serving as the precursor for $\rm HCN$ remains stable and is consistent with previous models that showed that $\rm HCN$ production depends most strongly on the carbon content of the atmosphere and is not strongly dependent on $\rm N_2$ or $\rm H_2$ abundances \citep{chameides+1981}. As we mentioned in Sect.~\ref{subsec:cantera}, the present atmosphere of Titan has primarily $\rm N_2$ which is thought to be generated from primordial ammonia \citep[e.g., ][]{ishimaru+2011,sekine+2011}. These results suggest that $\rm N_2$ is less affected by meteoroid impacts on the atmosphere, which is consistent with the fact that Titan has retained a large amount of $\rm N_2$ to the present day. 

\begin{figure}[htbp]
\centering
   {\includegraphics[width=9cm]{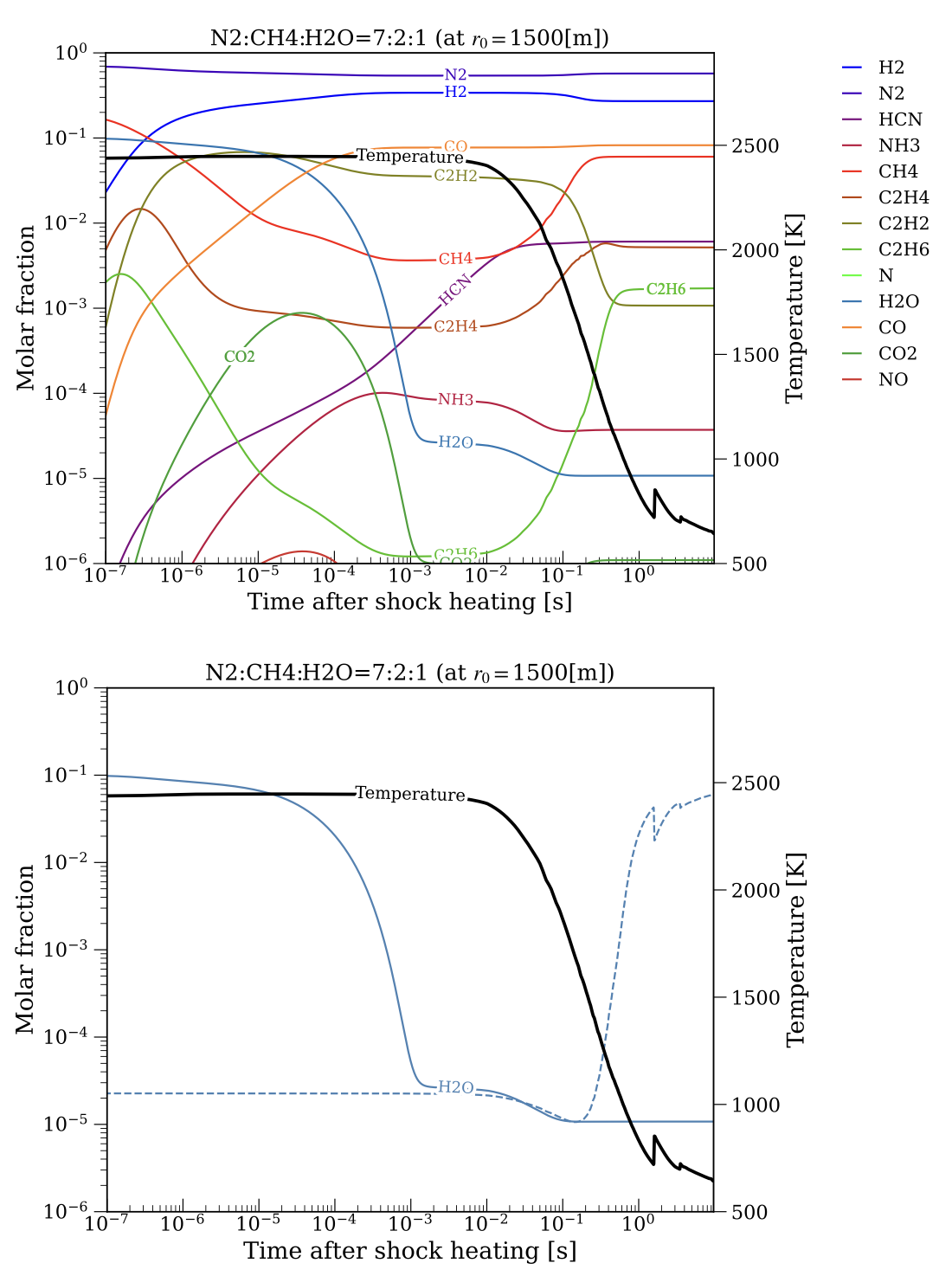}}
\caption{{Chemistry of atmosphere with water content.} In both two panels, the initial composition of atmosphere before shock heating includes 70\% of $\rm N_2$, 20\% of $\rm CH_4$ and 10\% of $\rm H_2O$. The upper panel shows the time evolution of chemical species with O-bearing molecules. The bottom panel indicates the comparison of the difference of $\rm H_2O$ between non-equilibrium simulation (solid line) and equilibrium simulation (dashed line). The initial position of the test particle is $r_0=1500$\,m. 
\label{fig:water}
}
\end{figure}

Observations by Herschel have confirmed the presence of $\rm H_2O$ vapor in Titan's atmosphere, with an estimated abundance of approximately $~\sim10^{-10}$\,molar fraction \citep[e.g.,][]{coustenis+1998,moreno+2012,horst2017}. 
Here, we perform chemistry simulations of an atmosphere that includes water as part of the initial composition. Figure~\ref{fig:water} shows a simulation result of chemical evolution of an atmosphere containing water vapor, which can be compared with the equilibrium case shown in Fig.~\ref{fig:eq_non1}. We note that, as in the previous cases, various molecules are actually produced; however, for clarity, the number of chemical species shown in the figure has been reduced. Water vapor decomposes due to shock heating, rapidly decreasing in abundance (blue-solid line) as it approaches its equilibrium composition (dashed line). However, once cooling begins ($t\gtrsim10^{-2}\,\rm s$), the water vapor amount deviates significantly from the equilibrium prediction, and eventually chemical quenching occurs without formation. As shown in Fig.~\ref{fig:water}, a single comet entry reduces the water vapor abundance by approximately four orders of magnitude, and oxygen is preferentially converted into CO gas. Therefore, even if Titan had acquired water in the past, its present limited water content could have been easily achieved. 

\subsubsection{Comparison with laser experiments}\label{subsubsec:experiment}
Some literature has reported chemical production using laser-induced plasma experiments to simulate meteor-driven chemistry in Titan's atmosphere \citep[e.g.,][]{flowers+2023}. {Although meteoroid entry is a far more energetic and large-scale phenomenon than can be reproduced in laboratory settings, laser-induced plasmas provide one of the few accessible analogs for studying high-temperature chemistry under controlled conditions. Therefore, in this study, we use the laser experiment as a representative reference for comparison.} Here, we compare our simulation results with these experimental findings. \citet{scattergood+1989} investigated chemical yields from laser-induced plasmas in a gas mixture composed of 90\% $\rm N_2$ and 10\% $\rm CH_4$, reporting production rates of various molecules as a function of the deposited energy. 

\begin{figure}[htbp]
\centering
   {\includegraphics[width=9cm]{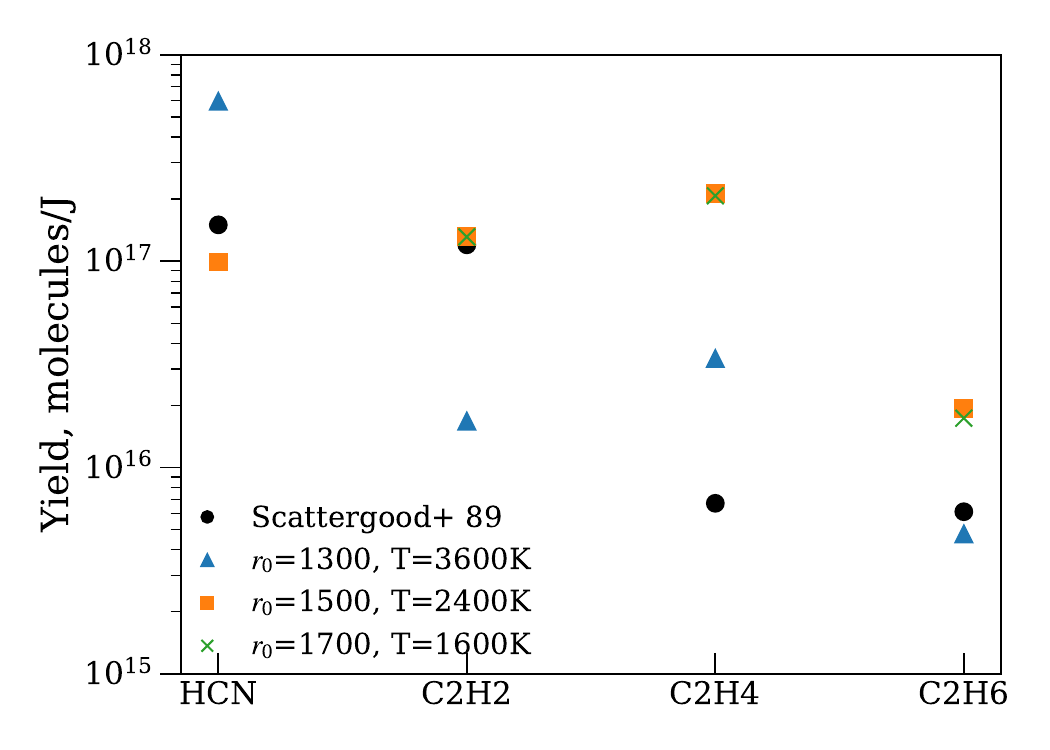}}
\caption{{Chemical production rates from our simulations and laser experiments by \citet{scattergood+1989}.} The black circles represent the experimental results. Colored symbols show our simulation results at three different radial distances ($r_0$) from the projectile, each corresponding to a different shock temperature ($T$). The horizontal axis lists molecular species (HCN, C\textsubscript{2}H\textsubscript{2}, C\textsubscript{2}H\textsubscript{4}, C\textsubscript{2}H\textsubscript{6}), and the vertical axis shows chemical yields in units of molecules per joule, which means that, for each species, the production rates are normalized by the total energy deposited at each location.
\label{fig:experiment}
}
\end{figure}

Figure~\ref{fig:experiment} presents a comparison between our simulations and the experimental results from \citet{scattergood+1989}. The amount of energy available for atmospheric alteration depends on the distance from the projectile, and we consider three representative radial distances, $r_0$. Each data point shows the cumulative chemical yields  of $\rm HCN,\ C_2H_2,\ C_2H_4$ and $\rm C_2H_6$, plotted against the energy deposited at each radial location. {These molecules are not only abundant in Titan's atmosphere--as confirmed by Cassini and mass spectrometer data \citep[e.g.,][]{vinatier2007} --but also play central roles in its organic chemistry. } For example, the blue triangles represent the production rates integrated along a vertical column at $r_0=1300$ m, where the shock temperature reaches 3600 K. {The bow shock generated by vertical entry is cylindrically symmetric. The yields at each $r_0$ are calculated from a cylindrical shell with a radius of $r_0$, a height of 100\,km, and a radial thickness of 20\,m which corresponds to one computational cell (Sect.~\ref{subsec:athena}).} For HCN, there is no green cross mark $(r_0,T)=(1700\,\rm m,\,1600\,K)$ because 1600\,K of shock temperature is not high enough to destroy the $\rm N_2$ bond to produce N-bearing components (see Fig.~\ref{fig:1700}). Our results indicate a general trend: lower shock temperatures tend to favor the formation of heavier molecules. This is consistent with the fact that larger organic compounds are more thermally fragile at higher temperatures. Although the bow-shock-induced chemistry investigated in this study differs significantly from the experimental setup in terms of post-shock expansion and chemical quenching timescales and temperatures, the chemical yields show good agreement with the experimental results (Fig.~\ref{fig:experiment}). 

\subsubsection{Integrated contribution by meteor entry}\label{subsubsec:hcn}

\begin{table*}[ht]
    \centering
    \begin{tabular}{c||c|cc}
        \hline \hline
        Species & Yields by the single entry [mol]\footnote{Total yields by the single entry-event of 1\,km sized body with the entry velocity $v_{\rm ent}=10\,\rm km/s$ (Sect.~\ref{subsec:athena})} & Present composition (molar fraction)\footnote{Chemical composition of Titan's neutral atmosphere reported by the Cassini measurement \citep{horst2017}} & Estimated abundance [mol]\\
        \hline
        HCN & 1.5$\times10^{11}$ & 6.7$\times10^{-8}$ & 1.4$\times10^{13}$ \\
        C\textsubscript{2}H\textsubscript{2} & 1.8$\times10^{10}$ & 3.0$\times10^{-6}$ & 6.3$\times10^{14}$ \\
        C\textsubscript{2}H\textsubscript{4} &  3.1$\times10^{10}$ & 1.2$\times10^{-7}$ & 2.5$\times10^{13}$ \\
        C\textsubscript{2}H\textsubscript{6} &  3.2$\times10^{9}$ & 7.3$\times10^{-6}$ & 1.5$\times10^{15}$ \\
        \hline \hline
    \end{tabular}
    \caption{Chemical yields and composition of Titan's atmosphere. The yields represents the cumulative productions of chemical species across the entire region. The present abundances are derived from the Cassini data. }
    \label{tab:yields}
\end{table*}

{
Chemical species are produced from the cylindrical region heated by the bow-shock wave, and its radius is $r\simeq 3R_{\rm com}$ (discussed below in Sect.~\ref{subsec:vs_impact}). We estimate the integrated production accumulated in the cylinder, which is shown in Table~\ref{tab:yields}, along with the present abundances in Titan's atmosphere reported by the Cassini observations. We note that while Fig.~\ref{fig:experiment} discussed the spatial dependence of chemical production, the total yields in Table~\ref{tab:yields} represent the integrated molecular production across the entire region heated by a single entry event of a cometary body that has the radius of $R_{\rm com}=1\,\rm km$ and the entry velocity of $v_{\rm ent}=10\,$km/s. }

{
The molar fractions shown in Table~\ref{tab:yields} represent the chemical composition of Titan's atmosphere at an altitude of $z \sim 100$\,km. While these values are used as representative in our analysis, several measurements have reported that atmospheric composition varies with altitude. For example, the molar fraction of HCN has been observed to increase with altitude and can reach values as high as $\sim 10^{-6}$ at $z \sim 400$\,km \citep[e.g.,][]{vinatier+10}, which is higher than the value listed in Table~\ref{tab:yields}. However, considering Titan’s atmospheric scale height of approximately $H = 20$\,km, most of the atmospheric mass resides at lower altitudes—for instance, $\rho(z=400\,\mathrm{km}) \lesssim 10^{-4} \rho_0$ \citep{fulchignoni+2005}. Therefore, for the purpose of estimating the total mass of each species in the atmosphere, using values from lower altitudes, such as those at $z = 100$\,km, provides a reasonable approximation. Then, based on the observed molar fractions, we estimate the current atmospheric abundances by assuming a molar mass of $M_{\rm mol}=28\,\rm g/mol$ and a total atmospheric mass of $M_{\rm atm,Titan}=9.0\times10^{18}\,\rm kg$\footnote{Estimated using $M_{\rm atm,Titan}=4\pi P_0R_{\rm T}^2g^{-1}$ where $P_0=1.5\,$bars, $g=1.3\rm\,m/s^2$ and $R_{\rm T}=2.6\times10^3\,$km}.}

{
By comparing the simulated chemical yields with the present-day abundances in Tab.~\ref{tab:yields}, we estimate the required accretion frequency to reproduce each observed species. HCN, a molecule regarded as a key precursor for biomolecules \citep[e.g.,][]{pearce+20}, can be reproduced by only about 100 entry events with $v_{\rm ent}=10\,$km/s and $R_{\rm com}=1$\,km (Tab.~\ref{tab:yields}). Given our assumed meteoroid mass of $M_{\rm met}=4\times10^{12}$\,kg, a hundred events correspond to merely $\sim10^{-4}$ of Titan's atmospheric mass, indicating that meteoroid entry events could have played a key role in HCN synthesis. Furthermore, as discussed in Sect.~\ref{subsubsuc:overview}, HCN production by the meteoroid event is largely independent of the initial chemical fraction in the pre-entry atmosphere. This implies that even if Titan’s atmospheric composition varied throughout its evolutionary history, meteoroid entry events could have consistently contributed to HCN synthesis. In contrast, to reproduce the present abundance of $\rm C_2H_6$, required frequency is $\sim5\times10^5$ entry events corresponding to the accreted mass of $\sim2\times10^{18}$\,kg. This total mass is comparable to $M_{\rm atm,Titan}=9\times10^{18}$\,kg; therefore, the entry event ($v_{\rm ent}=10\,$km/s and $R_{\rm com}=1$\,km) is unlikely to be responsible for reproducing the present abundance of $\rm C_2H_6$. 
}

{
Since the chemical production caused by meteoroid entry, as considered in our study, does not show strong dependence on altitude (Appendix~\ref{appendix3}), it is unlikely to be responsible for the observed altitude-dependent composition variations in Titan's atmosphere \citep[e.g.,][]{vinatier+10}. Therefore, other processes such as photochemistry and galactic cosmic ray (GCR) chemistry are more likely to account for the observed vertical distribution of molecular species \citep[e.g.,][]{loison+15,willacy+16,vuitton+19,pearce+20}. }

{
In conclusion, meteoroid entry events could have a significant role in supplying species, such as HCN, in Titan's early atmosphere. According to \citet{artemieva+2005}, which modeled the impact flux onto Titan's surface, a few billion years ago the number of impact events of the type considered in this paper could have exceeded 1000. Therefore, meteoroid events likely contributed substantially to the early atmospheric composition of Titan. In contrast, the photochemical processes are expected to dominate HCN production in recent times, as the photochemical timescale in Titan's atmosphere has been estimated to be only a few thousands years \citep{hebrard+2012}.} 

{In this study, we explore meteoroid-induced chemistry in Titan's lower atmosphere below 100\,km. Our results demonstrate that such entry events play a key role in enriching HCN in Titan's early lower atmosphere through non-equilibrium shock-driven chemistry, which strongly enhances HCN production compared to equilibrium predictions (e.g., Fig.~\ref{fig:eq_non1}). Although our present work focuses on the atmospheric entry phase, these meteoroids would eventually impact Titan’s surface, resulting in vaporization and melting \citep{miyayama2024}. The impact-induced melt pools may interact with haze organics, which are ultimately sourced from atmospheric HCN, and provide aqueous environments conducive to prebiotic chemical synthesis \citep{pearce+2024}. Therefore, meteoroid entry not only shaped the chemical composition of Titan’s early atmosphere but also may have helped set the stage for prebiotic environments. The subsequent impact phase is likely to be an important contributor to prebiotic chemistry and will be investigated in future work.
}

\subsubsection{Dominant reactions and rate coefficients}\label{subsubsec:rate_sensitivity}
{
We discuss the uncertainties associated with the rate coefficients used in the chemical kinetics network of this study, which is based on GRI-Mech 3.0. While this mechanism has been optimized for natural gas combustion, especially methane, it includes only limited C3 chemistry, making its predictions less reliable for heavier hydrocarbons. In high-temperature regimes, such as those induced by shock waves, rate coefficients are often poorly constrained due to limited experimental data. In some cases, uncertainties can span an order of magnitude \citep[e.g.,][]{pearce+20}. \citet{pearce+20} explored the effect of uncertainties on kinetic rate coefficients on the chemistry of the upper atmosphere of Titan, including the production of HCN at altitudes from $\sim100$ km to 1200\,km. We note that temperatures and pressures in these regions are considerably lower than those produced by the meteor entry process we explore here. The aerodynamical chemistry we explore here also takes place below 100\,km, where few observational constraints on Titan’s atmospheric chemistry are available. The flux of UV radiation that drives photochemistry is extremely reduced at altitudes below 100\,km, with fluxes of radiation below 190\,nm dropping to effectively zero at lower altitudes \citep{krasnopolsky2009}. Production via aerodynamical chemistry in a meteor-generated shock wave as we have discussed here may be a plausible source. } 

{To assess how such uncertainties may influence our results, we examine the dominant reaction pathways contributing to HCN production. Based on our simulation results, we identify the most significant reactions for HCN formation. As discussed above (e.g., Fig.~\ref{fig:distance}), chemical yields depend on the distance from the falling object. As a representative case, we analyze a test particle initially located at $(r,z) = (1500\,\mathrm{m}, -2000\,\mathrm{m})$, corresponding to the conditions shown in Fig.~\ref{fig:eq_non1}. As described in Sect.~\ref{subsubsec:experiment}, the contributions of HCN production are integrated over a cylindrical shell with a radius of $r_0 = 1500$\,m. Table~\ref{tab:ranking} lists the three most influential reactions along with their rate coefficients.}

{
The rate coefficients vary with temperature and pressure; the values listed in Table~\ref{tab:ranking} correspond to the shocked state with $T = 2447$\,K and $P = 5.5 \times 10^6$\,Pa, which occurs around $t \sim 10^{-7}$\,s in Fig.~\ref{fig:eq_non1}. In Tab.~\ref{tab:ranking}, we also list experimental data for the first reaction’s rate coefficient \citep{rodgers+96_k_coefficient,miller+97_k_coefficient}; however, no reliable data are available for the other two reactions at such high temperatures. According to these measurements, the uncertainty in the rate coefficient spans nearly two orders of magnitude. Since Cantera provides only the GRI-Mech mechanism for the C/H/N/O high-temperature chemistry, and no alternative network is readily available for comparison, we note that these three reactions are likely to have the greatest influence on HCN production under shock conditions. Therefore, the uncertainties in their rate coefficients could substantially influence the predicted chemical yields.}

\begin{table*}[]
    \centering
    \begin{tabular}{cccc}
        \hline\hline
        Contribution & Reaction equation & Rate coefficient, $k$\footnote{Second-order rate coefficients, $k$ in units of m$^3$\,molecule$^{-1}$\,s$^{-1}$, evaluated at $T=2447$\,K and $P=5.5\times10^6$\,Pa.} & Experimental, $k(2447\,\rm K)$\\
        \hline
        49\% & CH + N$_2$ $\rightarrow$ HCN + N & $8.0\times10^{-17}$ & $4.9-280\times10^{-14}$\footnote{\citet{rodgers+96_k_coefficient,miller+97_k_coefficient}}\\
        40\% & H$_2$CN $\rightarrow$ HCN + H & $1.7\times10^{-15}$ & \\
        9.4\% & CH$_3$ + N $\rightarrow$ HCN + H$_2$ & $2.0\times10^{-14}$ & \\
        \hline\hline
    \end{tabular}
    \caption{Dominant chemical reactions contributing to HCN formation in the shocked gas, along with their rate coefficients. The listed values are evaluated at $T=2447\,\rm K$ and $P=5.5\times10^6\,\rm Pa$, corresponding to the shocked state at $t\sim10^{-7}\,\rm s$ in Fig.~\ref{fig:eq_non1}. Experimental measurements are available for the first reaction and were taken from the NIST Chemical Kinetics Database; however, no reliable reference data exist for the other two reactions at this temperature.}
    \label{tab:ranking}
\end{table*}

\subsection{Aerodynamical chemistry vs impact induced chemistry}\label{subsec:vs_impact}
As mentioned in Sect.~\ref{sec:intro}, both aerodynamical heating during atmospheric entry and impact-induced vaporization contribute to planetary atmospheric evolution through shock heating and subsequent expansion. While previous studies have primarily focused on the role of impact-induced vaporization, we estimate and compare the contributions of these two processes from a mass perspective in this section.

\begin{figure}[htbp]
\centering
   {\includegraphics[width=9cm]{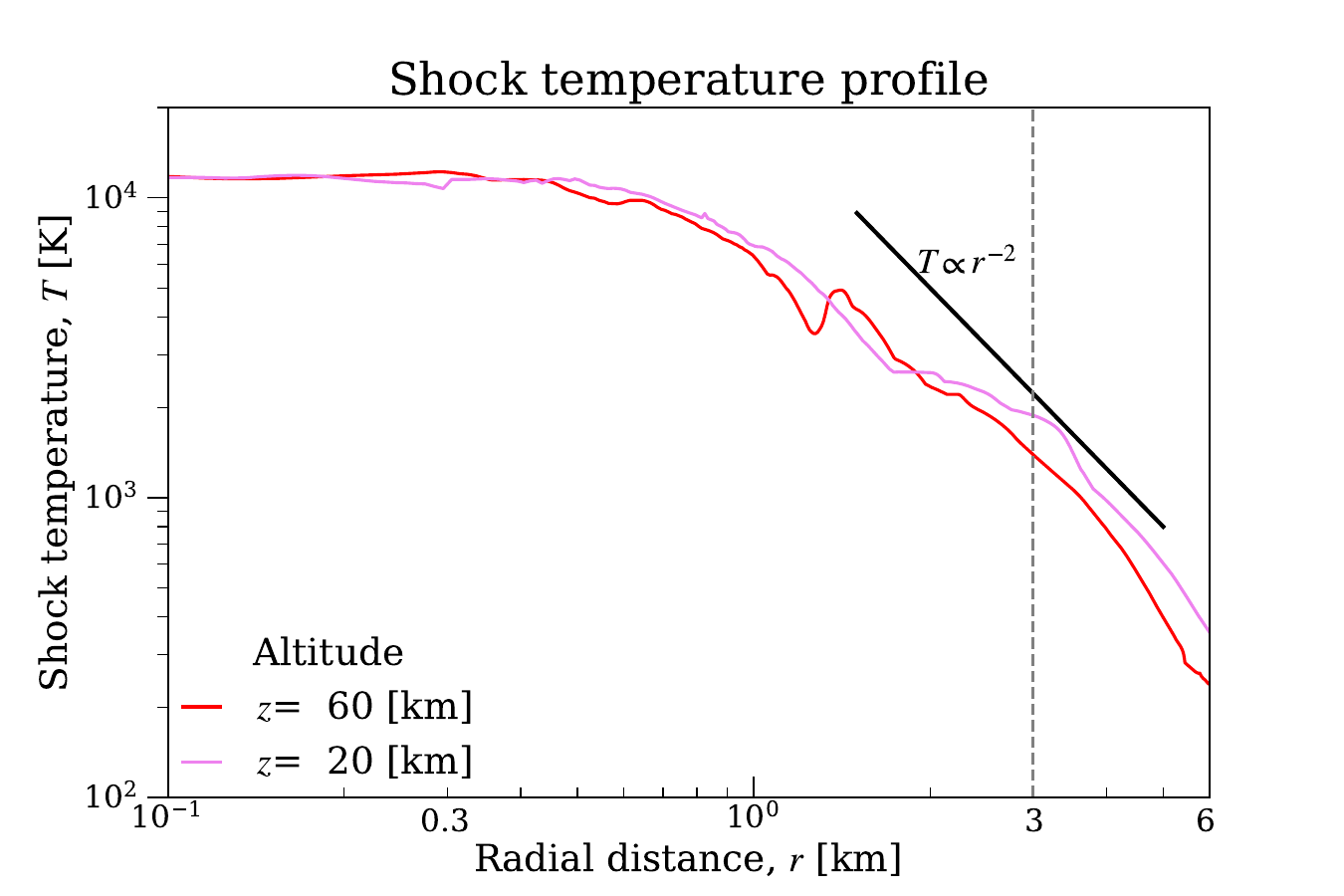}}
\caption{{Temperature profile due to bow shock heating for an impactor with $R_{\rm com}=1\rm \,km$.} Both lines show the maximum temperature at each radial distance. The color difference corresponds to altitude of the comet. Black line indicates a fitting line of $T\propto r^{-2}$. 
\label{fig:temp_prof}
}
\end{figure}

As shown in Fig.~\ref{fig:track}, shock temperature decreases with increasing distance from the comet, while the highest temperatures are experienced at the comet's leading edge due to the bow shock wave. Figure~\ref{fig:temp_prof} illustrates the shock temperature profiles as a function of radial distance. In our simulations, we assume that the atmosphere maintains a hydrostatic structure; therefore, density and pressure increase at lower altitudes. However, temperature is determined by the balance between density and pressure (Eq.~\ref{eq:pres_static}), leading to a shock temperature profile that remains independent of altitude (Fig.~\ref{fig:temp_prof}). According to Fig.~\ref{fig:temp_prof}, shock temperature remains nearly constant for $r\lesssim 10^0\,\rm km$. This behavior arises because the comet radius is 1 km where the flow directly collides with the comet's face. However, for $r\gtrsim 10^0\,\rm km$, the surrounding medium does not directly impact the object, leading to a decrease in temperature at greater distances, roughly following $T\propto r^{-2}$.

According to our results, a temperature exceeding 2000~K is required to trigger atmospheric evolution. Figure~\ref{fig:temp_prof}, which is derived from the 1\,km sized comet, indicates that the cross-sectional area contributing to atmospheric alteration is approximately $r\lesssim3\,\rm km$. The total mass of the atmosphere within this heated region can be estimated using the hydrostatic structure. The column density of the atmosphere, $\Sigma$, is given by
\begin{equation}
    \Sigma = \frac{P_0}{g},
\end{equation}
where $P_0$ is the surface pressure and $g$ is the gravitational acceleration. The total atmospheric mass, $M$, within a cylindrical volume of radius $r$ is given by
\begin{equation}
    M(r) = \pi r^2 \frac{P_0}{g}.
\end{equation}
Adopting $r=3\,$km, $P_0=1.3\times10^5\,$Pa, and $g=1.4\,\rm m/s^2$, the atmospheric mass that can contribute to chemical alteration is estimated as $M\sim4\times10^{12}\,\rm kg$.

\begin{figure}[htbp]
\centering
   {\includegraphics[width=9cm]{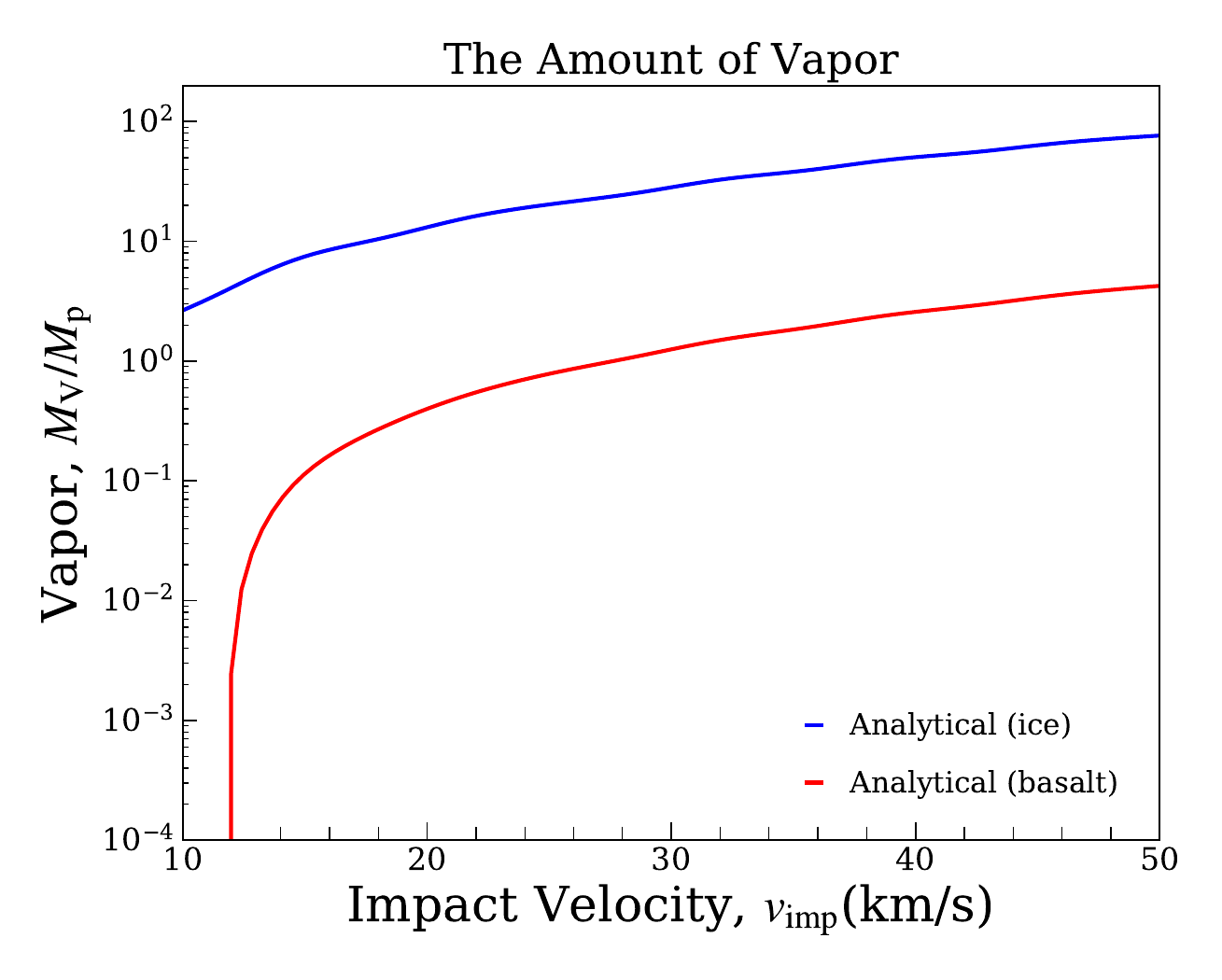}}
\caption{{Impact induced vapor as a function of impact velocity.} This figure is modified version of a figure from \citet{miyayama2024}. The amount of vapor $M_{\rm V}$ is scaled by a projectile mass $M_{\rm p}$. This assumes a vertical collision between same material, and each color indicates material types. 
\label{fig:vapmass}
}
\end{figure}

Our previous work \citep{miyayama2024}, on the other hand, investigated the mass contribution of impact-induced vapor shaping planetary atmospheres. Figure~\ref{fig:vapmass} presents the amount of vaporized mass as a function of impact velocity, assuming a vertical collision between identical materials. For a cometary impact on an icy surface, the mass of impact-induced vapor is given by $M \sim 3M_{\rm p}$, where $M_{\rm p}$ is the projectile mass, leading to an estimated $M\sim 4\times10^{12}\,$ kg. This suggests that the contribution of aerodynamical heating to atmospheric evolution is comparable to that of impact-induced vaporization for an icy surface.

However, according to the red line in Fig.~\ref{fig:vapmass}, an impact velocity of 10~km/s, which we have adopted as the entry velocity, is not sufficiently high to generate rocky vapor. This implies that if a planet is covered by a predominantly rocky surface, impact-induced vaporization becomes significantly less efficient due to the higher vaporization threshold. In such cases, aerodynamical heating during atmospheric entry may play a more dominant role in shaping planetary atmospheric evolution.

\section{Conclusion}\label{sec:conclusion}
The atmospheric entry of a meteoroid generates a bow shock wave, leading to intense atmospheric heating. In such high-temperature gas, chemical reaction rates increase significantly, and shock-induced chemistry can alter the atmospheric composition. The rapid chemical evolution continues until the cooling timescale due to expansion exceeds the reaction timescales, at which point chemical quenching occurs. Consequently, the final chemical composition is essentially determined by the shock heating and subsequent cooling processes. 

To investigate this process, we have simulated the aerodynamical flow of a meteoroid entry using the Athena++ code. For the meteoroid material, we adopted a simple equation of state (ideal-gas EoS) instead of a solid-phase EoS. This assumption is justified by the large density contrast between the meteoroid and the atmospheric gas, which typically exceeds $10^3$. Our simulations revealed the formation of a secondary shock wave behind the meteoroid as well as the primary bow shock. We also analyzed the thermodynamic evolution along a streamline of a test particle in this complex shock-driven flow and examined atmospheric mixing with meteoritic material by solving a diffusion equation. Furthermore, we evaluated the expansion cooling timescale, which governs chemical quenching, and demonstrated its dependence on meteoroid size and velocity. 

Using the test particle analysis, we coupled shock-induced chemistry with fluid dynamics. In the high-temperature region near the meteoroid, chemical species rapidly equilibrate due to the strong temperature dependence of reaction rates, and end up with the formation of small molecules. Further from the comet body, on the other hand, polyatomic species begin to form. Comparing our results between equilibrium and non-equilibrium chemistry models, we found that chemical quenching can occur before reaching equilibrium in regions where shock temperatures are not sufficiently high. The influence of the secondary shock wave on chemical evolution varies depending on the species, suggesting that different chemical pathways may be activated under varying shock conditions. 

We discussed the implications of meteoroid entry for Titan’s current atmosphere. Nitrogen, the dominant component of Titan’s atmosphere, is a highly stable species, consistent with its present-day abundance. Some previous studies have suggested the presence of water vapor on Titan. Our results indicate that meteoroid entry can efficiently remove {$\rm H_2O$ and largely enhance HCN}, a process that is not adequately captured by equilibrium chemistry models. We also evaluated the cumulative impact of meteoroid entry events on Titan’s atmosphere as a whole. By comparing the estimated chemical yields from entry-driven synthesis with the observed molecular abundances reported by the Cassini-Huygens mission, {we find that this process could have played a significant role in supplying certain key organic molecules, particularly HCN, during early Titan's history.} Our results suggest that even a modest number of meteoroid entry events—on the order of ~100 for kilometer-scale objects—could account for the current abundance of HCN, highlighting the importance of entry chemistry as a complementary mechanism to photochemistry and other proposed pathways. 

Moreover, we compared our simulation results with laboratory experiments using laser-induced plasma in a Titan-like atmosphere. The comparison shows that, despite the different physical conditions, the overall trends and yields of key organic species such as HCN and C\textsubscript{2}H\textsubscript{2} agree well, supporting the validity of our approach.

Also, we compared the contributions of meteoroid atmospheric entry and impact-generated vapor to atmospheric chemical evolution. We analyzed the temperature distribution formed by the bow shock during atmospheric entry and identified that the key region for chemical evolution is where temperatures exceed 2000\,K. Using the temperature distribution, we estimated the total mass of the atmosphere that can undergo chemical alteration. Furthermore, by comparing our estimates of atmospheric evolution through meteoroid entry with the vapor mass generated by impact events from our previous study \citep{miyayama2024}, we assessed the relative importance of these two mechanisms. Our results suggest that in planetary environments covered predominantly by a rocky surface, the contribution from impact-generated vapor may be less significant, and instead, the atmospheric entry process could play a more dominant role in shaping atmospheric composition.  

\begin{acknowledgements}
We thank the developers of Athena++ for providing the code used in our simulations. All simulations for this project were performed on the Sherlock cluster. We would like to thank Stanford University and the Stanford Research Computing Center for providing computational resources and support that contributed to these research results. We used the iSALE code for validation of our calculations. We thank iSALE developers, including G. Collins, K. Wünnemann, B. Ivanov, J. Melosh, and D. Elbeshausen and Tom Davison for development of pySALEplot. RM has been supported by Interdisciplinary Frontier Next-Generation Researcher Program of the Tokai Higher Education and Research System and the Japan Society for the Promotion of Science (JSPS) KAKENHI grant number JP22K21344. AZ has been supported by the NASA FINESST Fellowship 80NSSC22K1323.
\end{acknowledgements}

\bibliographystyle{aasjournal} 

\begin{appendix}
\section{Estimation of timescales}\label{appendix2}
As described in Sect.~\ref{subsec:athena}, we solve ideal fluid equations to simulate the atmospheric entry, which do not include dissipation processes by heat conduction and radiation. This is because the cooling by adiabatic expansion is the dominant process, making those dissipation physics minor. Here we estimate the timescales of both processes. In the following section, $L\, \rm [m]$ represents a characteristic length of the system. 

Heat conduction is simply modeled by the diffusion equation: 
\begin{eqnarray}
    \frac{\partial T}{\partial t}=\kappa\frac{\partial^2 T}{\partial x^2},
    \label{eq:heatcond}
\end{eqnarray}
where $\kappa$ is the material thermal diffusivity in $\rm m^2\,s^{-1}$, which is typically $10^{-6}\,\rm m^2\,s^{-1}$ for a solid material. Therefore, the cooling timescale of heat conduction $\tau_{\rm heat}$ is estimated as
\begin{eqnarray}
    \tau_{\rm heat} \sim L^2/\kappa=    10^6\,\left(\frac{L}{1\,\rm [m]}\right)^2
    \left(\frac{\kappa}{10^{-6}\,[\rm m^2\,s^{-1}]}\right)^{-1}    \,[\rm s]. 
    \label{eq:heat_time}
\end{eqnarray}
For radiative cooling, assuming the radiative flux is given by Stefan-Boltzmann's law, temperature changes through spherical black body radiation are described as
\begin{eqnarray}
    \frac{4\pi}{3}L^3\rho C_{\rm p}\frac{dT}{dt}=-4\pi L^2\sigma T^4,
\end{eqnarray}
where $\rho,C_{\rm p},\sigma$ are density, heat capacity of gas, and Stefan-Boltzmann constant, respectively. Therefore the timescale for the radiative process is estimated as
\begin{eqnarray}
    \tau_{\rm rad}& \sim& \frac{\rho C_{\rm p}L}{3\sigma T^3}\\&\sim& 10^2\,\left(\frac{\rho C_{\rm p}}{10^3\,\rm [J\,kg^{-3}K^{-1}]}\right)\,\left(\frac{L}{1\,\rm [m]}\right)
    \left(\frac{T}{10^3\,[\rm K]}\right)^{-3}    \,[\rm s]. \label{eq:rad_time}
\end{eqnarray}
As shown in Fig.~\ref{fig:track}, the length scale of bow shock flow is about a few $\rm km$, making $\tau_{\rm heat}\sim10^{12} ,\ \tau_{\rm rad}\sim 10^5$ seconds. Therefore, comparing the timescale of expansion cooling $\tau_{\rm cool}\sim10^{-2}\,\rm s$ (Fig.~\ref{fig:cooling}), the adiabatic expansion is dominant as a cooling process. 

\section{Validation of the ideal-gas EoS} \label{appendix1}
Although, in reality, meteoroids are solid, we simply use ideal gas EoS instead of a solid EoS because Athena++ \citep{stone+2020} does not support solid phase. Here, we validate the ideal-gas EoS assumption using iSALE code \citep{Amsden_etal_1980,Ivanov_etal_1997,2006Icar..180..514W} which supports solid models and is frequently used in planetary-impact simulations. 

To validate our assumption, we simulate atmospheric entry based on the both EoS models, ideal-gas EoS (Eq.~\ref{eq:eos}) and Tillotson~EoS model \citep{1962geat.rept.3216T} by using iSALE code. Tillotosn~EOS is widely used in planetary-impact simulations and is described as: 
\begin{eqnarray}
    P=(\gamma-1)\rho E + A(\rho/\rho_0-1) + B(\rho/\rho_0-1)^2.
    \label{eq:tillo}
\end{eqnarray}
The second and third terms with only dependence on density are specific to solid bodies which are called the cold terms \citep[e.g., ][]{2007MAPS...42.2079M}, while the first term is common term in gas EoS (Eq.~\ref{eq:eos}). 

\begin{figure}[htbp]
\centering
   {\includegraphics[width=8cm]{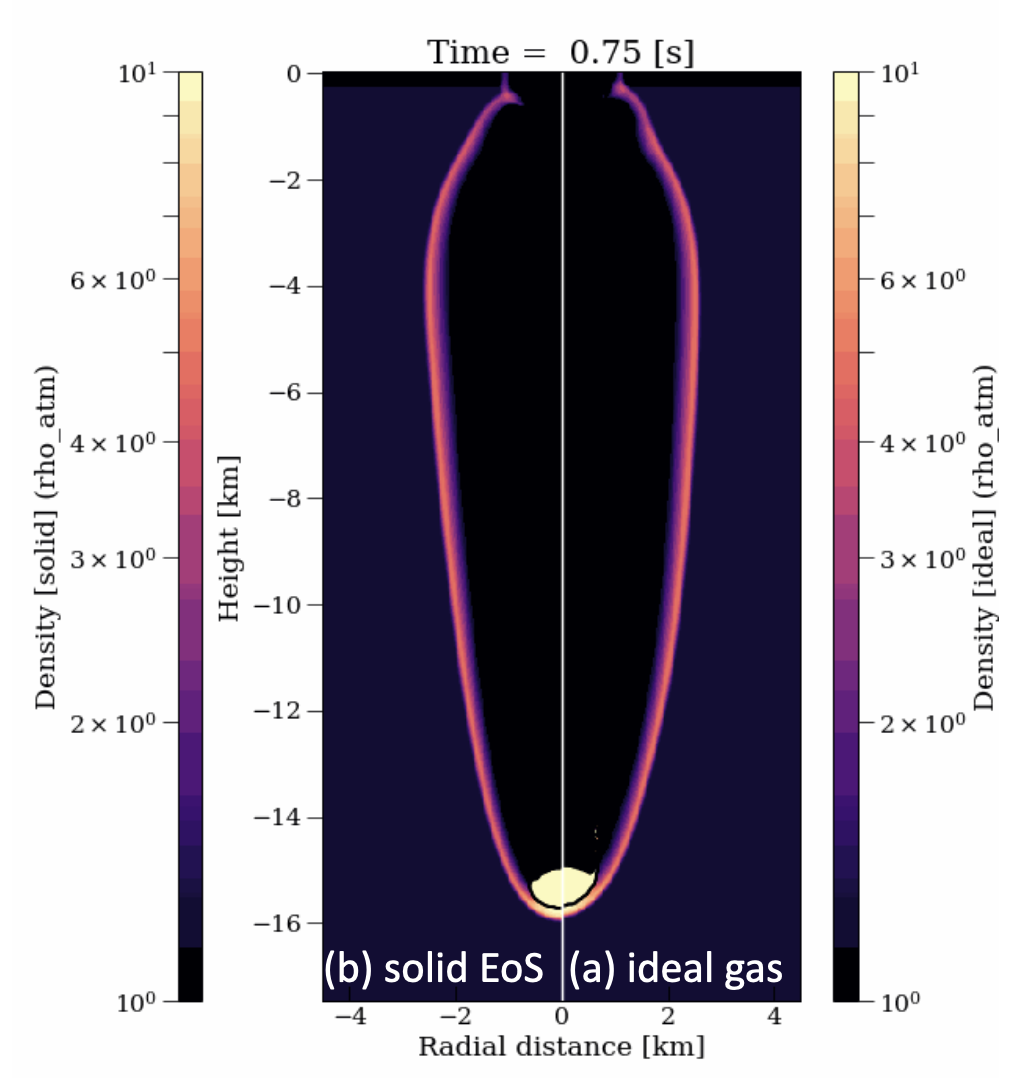}}
\caption{{Two density profiles by solid and ideal-gas EoS.} Right and left panels show density color maps by using different EoS-model--the ideal-gas EoS (a) and the solid EoS (b). Meteoroid bodies move with an entry velocity $v_{\rm ent}=20\,\rm km/s$ to bottom side of medium with homogeneous density, $\rho_{\rm atm}=1.2\,\rm kg/m^3$. The density of the entry meteoroid is $10^3\rm \, kg/m^3$, and color map of density is scaled by $\rho_{\rm atm}$. 
\label{fig:isale}
}
\end{figure}

Figure~\ref{fig:isale} shows density profile at $t=0.75\,\rm s$ obtained from iSALE simulation with two different EoS-models, simulating atmospheric entry with an entry velocity of $20\,\rm km/s$. We adopt the ideal-gas EoS (Eq.~\ref{eq:eos}) with $\gamma=1.4,\,\rho_0=10^3\,\rm kg/m^3$ for the meteoroid in the right panel, while the solid EoS with same $\gamma$ and $\rho_0$ and with $A=B=10^{10}\,[\rm Pa]$ is used in the left panel. As we discussed in Sect.~\ref{sec:aerodynamics}, bow-shock flow is generated in both panels in Fig.~\ref{fig:isale}, and those flows look similar. Although the difference of EoS-models slightly influences the shape of meteoroids, Fig.~\ref{fig:isale} shows that ideal-gas EoS is sufficient enough to investigate the bow-shock flow. It has been investigated that the iSALE code tends to underestimate temperature (Fig. 5 in {\citealp{miyayama2024}}), which is crucial for chemical reactions. Therefore, we use the Athena++ code \citep{stone+2020}, which provides a more accurate temperature evaluation, for our simulations. Consequently, we assess the validity of the ideal gas equation of state for atmospheric entry. 

\section{Altitude dependence of the chemistry} \label{appendix3}
We present the atmospheric evolution results at an altitude of 100 km as representative (Sect.~\ref{subsec:chem_result} and \ref{sec:discussion}), but as explained in Sect.~\ref{subsec:vs_impact}, atmospheric pressure and density vary with altitude, while temperature is independent. Here we mention an influence of these variations on the chemistry. 

\begin{figure*}[htbp]
\centering
    \begin{subfigure}[t]{0.45\textwidth}
      \caption{Altitude: $z_0$=60\,km} 
      \includegraphics[width=\textwidth]{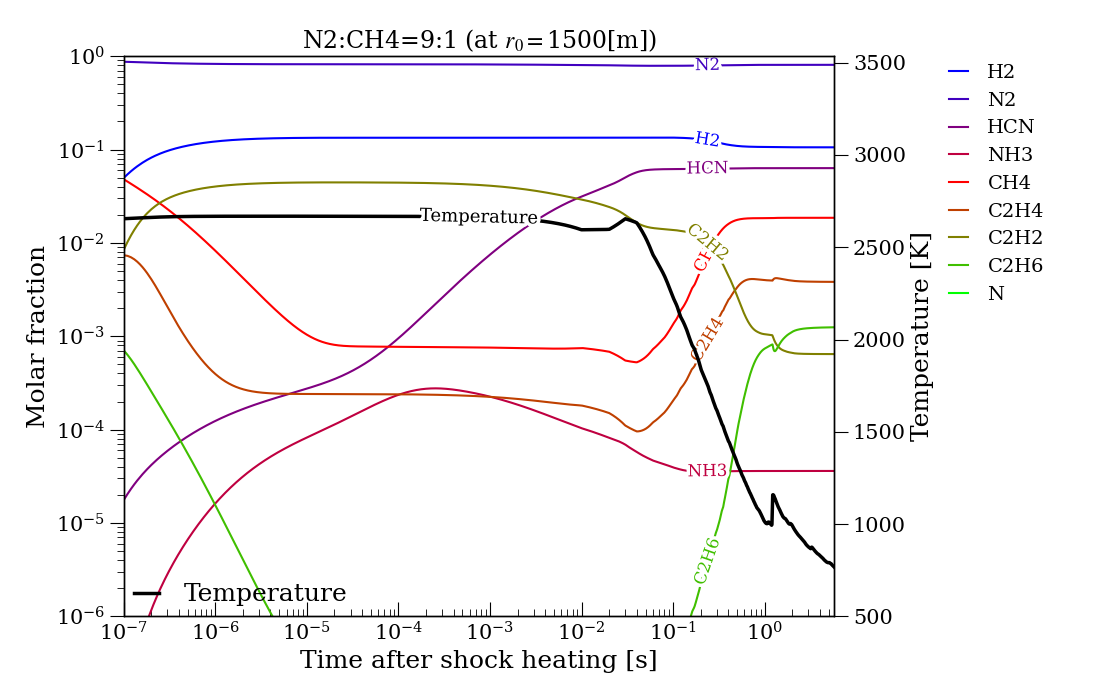}\label{subfig:z40}
    \end{subfigure}
    \hfill
    \begin{subfigure}[t]{0.45\textwidth}
      \caption{Altitude: $z_0$=40\,km}
      \includegraphics[width=\textwidth]{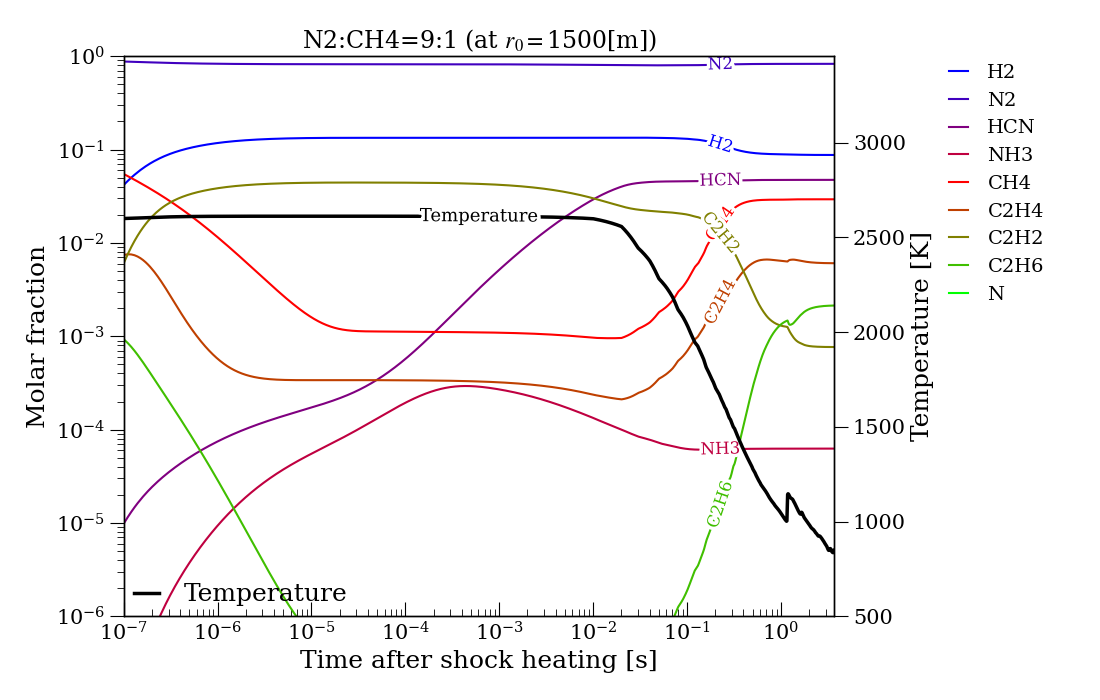}\label{subfig:z60}
    \end{subfigure}
   \caption{Altitude dependence of the chemistry. Fig.~(a) indicates the results for a test particle initially located at $(r_0,z_0)=\rm(1500\,m,60\,km)$, and Fig.~(b) for one initially located at $(r_0,z_0)=\rm(1500\,m,40\,km)$. \label{fig:altitude}
}
\end{figure*}

Figure~\ref{fig:altitude} shows the chemistry results from two different altitude case $z=60$\,km and $z=40$\,km. We set the atmospheric scale height to $H=20\,$km, which corresponds to the altitude where pressure and density become $1/e$ of their values at the ground surface; therefore, the pressure in Fig.~\ref{subfig:z60}.1(b) is $e$ times larger than that in Fig.~\ref{subfig:z40}.1(a). As shown in Figs.~\ref{fig:distance}, the chemistry is highly sensitive to temperature. In contrast, comparing two panels in Fig.~\ref{fig:altitude}, the impact of pressure and density variation, which correspond to changes in altitude, on the chemistry is minor.

\end{appendix}



\end{document}